\documentclass[aps,pra,superscriptaddress,reprint, twocolumn, amsmath, amssymb, a4, notitlepage]{revtex4-1}
\usepackage{graphicx}
\usepackage{amsmath,amssymb}
\usepackage{dcolumn}
\usepackage{mathrsfs}
\usepackage{physics}
\usepackage[utf8]{inputenc}
\usepackage{float}
\usepackage{color}

\begin{document}
\title{Non-Markovian perturbation theories for phonon effects in strong-coupling cavity quantum electrodynamics}
\date{\today}

\author{Matias Bundgaard-Nielsen}
\affiliation{Department of Photonics Engineering, DTU Fotonik, Technical University of Denmark, Building 343, 2800 Kongens Lyngby, Denmark}
\affiliation{NanoPhoton-Center for Nanophotonics, Technical University of Denmark, Building 343, 2800 Kongens Lyngby, Denmark}
\author{Jesper M\o rk}
\affiliation{Department of Photonics Engineering, DTU Fotonik, Technical University of Denmark, Building 343, 2800 Kongens Lyngby, Denmark}
\affiliation{NanoPhoton-Center for Nanophotonics, Technical University of Denmark, Building 343, 2800 Kongens Lyngby, Denmark}
\author{Emil Vosmar Denning}
\affiliation{Department of Photonics Engineering, DTU Fotonik, Technical University of Denmark, Building 343, 2800 Kongens Lyngby, Denmark}
\affiliation{NanoPhoton-Center for Nanophotonics, Technical University of Denmark, Building 343, 2800 Kongens Lyngby, Denmark}
\affiliation{Nichtlineare Optik und Quantenelektronik, Institut f\"ur Theoretische Physik, Technische Universit\"at Berlin, Berlin, Germany}

\begin{abstract}
Phonon interactions are inevitable in cavity quantum electrodynamical systems based on solid-state emitters or fluorescent molecules, where vibrations of the lattice or chemical bonds couple to the electronic degrees of freedom. Due to the non-Markovian response of the vibrational environment, it remains a significant theoretical challenge to describe such effects in a computationally efficient manner. This is particularly pronounced when the emitter-cavity coupling is comparable to or larger than the typical phonon energy range, and polariton formation coincides with vibrational dressing of the optical transitions. In this Article, we consider four non-Markovian perturbative master equation approaches to describe such dynamics over a broad range of light-matter coupling strengths and compare them to numerically exact reference calculations using a tensor network. The master equations are derived using different basis transformations and a perturbative expansion in the new basis is subsequently introduced and analyzed. We find that two approaches are particularly successful and robust. The first of these is suggested and developed in this Article and is based on a vibrational dressing of the exciton-cavity polaritons. This enables the description of distinct phonon-polariton sidebands that appear when the polariton splitting exceeds the typical phonon frequency scale in the environment. The second approach is based on a variationally optimized polaronic vibrational dressing of the electronic state. Both of these approaches demonstrate good qualitative and quantitative agreement with reference calculations of the emission spectrum and are numerically robust, even at elevated temperatures, where the thermal phonon population is significant.
\end{abstract}
\maketitle

\section{Introduction}
\label{sec:intro}
Quantum technology relies on the generation and processing of fragile quantum mechanical states \cite{O_brian_semiconductor_2009}. Interactions with the surrounding environment inevitably destroy the coherence and entanglement of these states. In solid-state quantum devices, one such interaction is with lattice vibrations or phonons, and cannot be avoided even at absolute zero temperature \cite{simultanoeusefficiency}. As an example, it has been shown, that scattering with phonons imposes a fundamental trade-off between the indistinguishability and the efficiency of single-photon sources \cite{simultanoeusefficiency}. These interactions can have a highly non-Markovian \cite{non_markovian_QD} and complex nature which leads to persistent optical emission features such as broad spectral sidebands~\cite{besombes2001acoustic}, incoherent scattering \cite{lightscattering,koong2019fundamental} and damping of coherent Rabi oscillations~\cite{forstner2003phonon,ramsay2010damping,ramsay2010phonon,ramsay2011effect}.

In the regime where the light-matter coupling exceeds the typical vibrational frequency of the environment, the nature of the electron-phonon coupling drastically changes, as compared to the weak light-matter coupling regime. Here, the strong emitter--cavity interaction can significantly influence, and in some cases decouple, the vibrational dynamics~\cite{Feist_2016_Supressing_photochemical,Frank_2016_cavity_controllerd,galego2015cavity}. In this regime, the optical emission spectrum exhibits two distinct polariton peaks, which are both dressed with a distinct vibrational sideband~\cite{Denning_2020_phonon_decoupling,hughes2021resonant}. On the other hand, resonant phonon-induced transitions between the polariton states is a dominating effect in the intermediate regime, where the polariton splitting is comparable with the typical environmental phonon frequency. This effect is an important resource for polariton condensation and -lasing~\cite{doan2005condensation,mazza2013microscopic,kasprzak2008formation}.

Several recent developments in a broad range of nanophotonics platforms have led to cavity quantum electrodynamical systems with very strong light-matter coupling, which in many cases is comparable to or exceeds the typical frequencies of the phonons that couple to the electronic degrees of freedom~\cite{chikkaraddy2016single,wang2016coherent,liu2017strong,kleemann2017strong,stuhrenberg2018strong,han2018rabi,geisler2019single,qin2020revealing,gross2018near,self_similar_Choi_2017,Shuren_exp_sub_2018,jesm_max_qual}. These experimental developments call for theoretical tools that are accurate and stable in this regime of cavity quantum electrodynamics.
A theoretical description that is able to describe the non-Markovian phonon response over the full range of light-matter coupling strengths is, however, a difficult task, and considerable efforts have been devoted to developing non-perturbative and non-Markovian methods \cite{tens_prev_1,tens_prev_2,tens_prev_3,tens_prev_4,tens_prev_5}. On the one hand, non-perturbative numerical calculations of the dynamics and emission properties can provide results with high numerical precision, but are computationally expensive and often do not provide physical insight to the results. On the other hand, semi-analytic perturbative methods are computationally efficient and can in many cases yield improved physical understanding, e.g. through analytical results. An important challenge thus lies in identifying the most accurate perturbative method in a given situation and parameter regime.

In this paper, we have implemented a computationally efficient tensor network formulation, which allows us to calculate two-time averages to any desired numerical precision \cite{Emil_tensor1,Emil_tensor2,Denning_2020_phonon_decoupling}. These calculations are then used as a benchmark to evaluate the accuracy of various less complex and less computationally demanding and more physically intuitive perturbative master equations across a large range of light--matter coupling strengths. By encoding electron-phonon or polariton-phonon correlations differently into the basis states prior to a perturbative expansion, the master equations are able to capture different non-Markovian effects. Specifically, we compare the performance of four master equation approaches: A new so-called polariton-polaron master equation, a variational polaron master equation, a standard polaron master equation, and a weak phonon master equation.

As a test system, we use a nanocavity containing a semiconductor quantum dot coupled to a continuum of longitudinal acoustic phonon modes of the host lattice \cite{phonon_1,phonon_2,phonon_3,phonon_4}. The primary quantity used for comparison of the methods is the optical emission spectrum, which relies on the calculation of the two-time correlation function of the cavity mode. Two-time averages are generally more sensitive to non-Markovian effects than one-time averages \cite{dara_2time_sens,pollock2018operational}, and are thus more challenging to correctly calculate with perturbative methods. Thus, by using the emission spectrum for comparison, the ability of the methods to capture the full non-Markovian phonon response is more clearly exposed as compared to evaluation of one-time averages.

The polariton-polaron approach is found to be the most accurate method in the strong-coupling regime where phonons manifest themselves as sidebands on the polariton peaks \cite{Denning_2020_phonon_decoupling,hughes2021resonant}. The variational approach is on the other hand found to be precise in the Purcell regime where the zero-phonon line acquires a phonon-sideband \cite{simultanoeusefficiency,phononeffects_emil}.

The Article is organized as follows: In section II, we introduce the model used to study phonon-coupled cavity quantum electrodynamics. In section III, we derive the perturbative master equations and introduce the various transformations that lead to these master equations. In section IV, we discuss possible ways of calculating the emission spectrum using the derived master equations and discuss how the different transformations enable the inclusion of phonon memory effects. In section V, we benchmark the master equations with a numerically convergent tensor network in the strong light-matter coupling regime and in the Purcell regime. In section VI, we analyze the strength of the perturbation in the different approaches. In Section VII, we discuss non-Markovian effects in one- and two-time averages followed up by a conclusion in section VIII.



%


\section{Model}
\label{sec:model}
The model we consider in this paper consists of a localized exciton state $\ket{X}$ that couples with a cavity mode with annihilation (creation) operator $a$ ($a^\dagger$) through the Jaynes-Cummings model \cite{tens_prev_2,phononeffects_emil}:
\begin{align}
\begin{split}
H_{\rm S} &= \hbar\omega_{eg}\sigma^\dagger\sigma + \hbar\omega_{c}a^\dagger a + \hbar g(\sigma^\dagger a + a^\dagger \sigma) \label{eq:H_S}
\end{split}
\end{align}
where $\omega_{eg}$ is the exciton frequency, $\omega_c$ the cavity mode frequency, $g$ the light-matter coupling strength and $\sigma = \ket{g}\bra{X}$ the annihilation operator for the exciton, with $\ket{g}$ being the ground state. The exciton is furthermore coupled with a bath of phonons through the term \cite{tens_prev_2,phononeffects_emil,manyparticle}
\begin{align}
H_{\rm I} &= \sigma^\dagger\sigma\sum_\mathbf{k}\hbar g_\mathbf{k}(b_\mathbf{k}^\dagger + b_\mathbf{k})
 \label{eq:H_I}
\end{align}
where $g_\mathbf{k}$ denotes the coupling strength to the phonon mode with momentum $\mathbf{k}$ created by the operator $b_\mathbf{k}^\dagger$, see eg. \cite{Iles_Smith_2017_Limits_to_coherent}. The free energy of the phonons is given by \cite{tens_prev_2,phononeffects_emil,manyparticle}:
\begin{align}
    H_{\rm E} &= \sum_\mathbf{k} \hbar \nu_\mathbf{k} b_\mathbf{k}^\dagger b_\mathbf{k},
     \label{eq:H_E}
\end{align}
where $\nu_\textbf{k}$ is the frequency of the phonon mode $b_\mathbf{k}^\dagger\ket{0}$. The central quantity that characterizes the influence of the phonon environment on the dynamics of the exciton-cavity system is the spectral density, defined as
\begin{align}
    J(\nu) = \sum_\mathbf{k} \abs{g_\mathbf{k}}^2\delta(\nu-\nu_\mathbf{k}).
\end{align}
The spectral density generally depends on the shape of the exciton wavefunction and the nature of the phononic environment. In this paper, we consider as an example system a semiconductor quantum dot in a spherically harmonic confinement potential, coupled to longitudinal acoustic phonons. In this case, the spectral density can be approximated as~\cite{kaer2012microscopic,phononeffects_emil} $J(\nu) = \alpha \nu^3 \exp(-\nu^2/\nu_{\rm c}^2)$ where $\alpha$ and $\nu_{\rm c}$ are parameters that depend on the size of the quantum dot and the properties of the surrounding material. The parameter $\alpha$ describes the overall strength of the coupling and $\nu_c$ is a cutoff frequency, which sets the frequency scale around which the interaction with phonons is strongest.
In addition to the effects generated by the Hamiltonians in Eqs.~\eqref{eq:H_S}--\eqref{eq:H_E}, decay of the cavity mode with rate $\kappa$ is included in the dynamical evolution as a Markovian effect. In a similar fashion, temperature-dependent pure dephasing due to higher-order phonon scattering effects is also included. This will be further elaborated on in the following section, where the derivation of different perturbative master equations from this fundamental model is considered.

\section{Perturbative master equations}
\begin{figure}[t]
  \centering
  \def\svgwidth{\linewidth}
\begingroup%
  \makeatletter%
  \providecommand\color[2][]{%
    \errmessage{(Inkscape) Color is used for the text in Inkscape, but the package 'color.sty' is not loaded}%
    \renewcommand\color[2][]{}%
  }%
  \providecommand\transparent[1]{%
    \errmessage{(Inkscape) Transparency is used (non-zero) for the text in Inkscape, but the package 'transparent.sty' is not loaded}%
    \renewcommand\transparent[1]{}%
  }%
  \providecommand\rotatebox[2]{#2}%
  \newcommand*\fsize{\dimexpr\f@size pt\relax}%
  \newcommand*\lineheight[1]{\fontsize{\fsize}{#1\fsize}\selectfont}%
  \ifx\svgwidth\undefined%
    \setlength{\unitlength}{412.45301254bp}%
    \ifx\svgscale\undefined%
      \relax%
    \else%
      \setlength{\unitlength}{\unitlength * \real{\svgscale}}%
    \fi%
  \else%
    \setlength{\unitlength}{\svgwidth}%
  \fi%
  \global\let\svgwidth\undefined%
  \global\let\svgscale\undefined%
  \makeatother%
  \begin{picture}(1,0.88167591)%
    \lineheight{1}%
    \setlength\tabcolsep{0pt}%
    \put(0,0){\includegraphics[width=\unitlength,page=1]{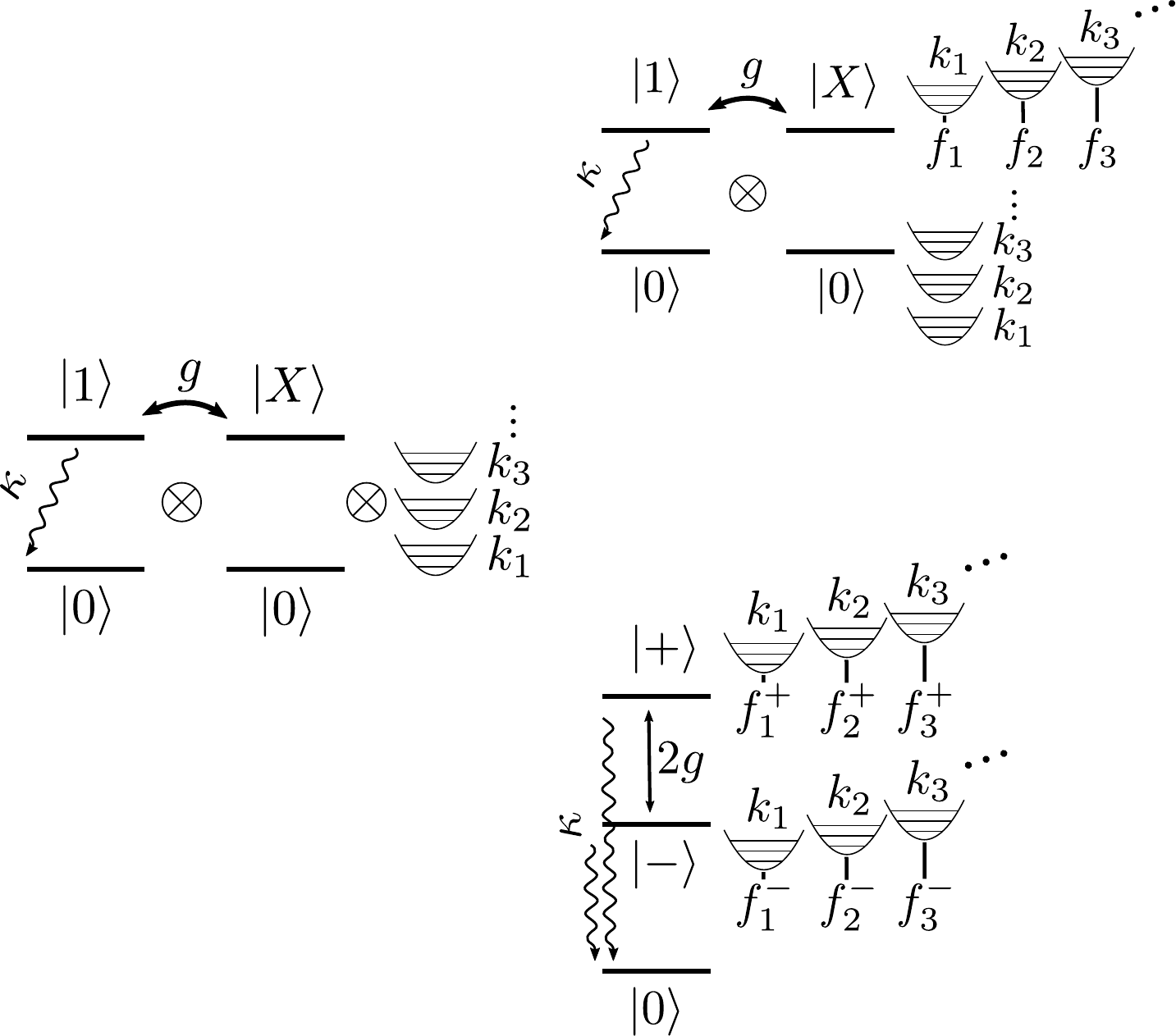}}%
    \put(0.11201429,0.12744673){\rotatebox{-0.17203575}{\makebox(0,0)[lt]{\lineheight{1.25}\smash{\begin{tabular}[t]{l}Polariton-Polaron\\transformation\end{tabular}}}}}%
    \put(0,0){\includegraphics[width=\unitlength,page=2]{System_scheme_phonons_v5.pdf}}%
    \put(0.11048692,0.80272302){\rotatebox{0.61006514}{\makebox(0,0)[lt]{\lineheight{1.25}\smash{\begin{tabular}[t]{l}Polaron-\\transformation\end{tabular}}}}}%
    \put(0,0){\includegraphics[width=\unitlength,page=3]{System_scheme_phonons_v5.pdf}}%
  \end{picture}%
\endgroup%

  \caption{A sketch of the system considered in this paper. Middle: A cavity mode, with decay rate $\kappa$ is coupled to a two-level emitter, which in turn is coupled to phonon modes $b_\textbf{k}$. Two transformations of the system are illustrated. In the Polaron transformation, the phonons dress the emitter and in the Polariton-polaron transformation, the phonons dress the coupled emitter-cavity polaritons. In both cases, the effect of the polaron transformation and the polariton-polaron transformation is shown. As depicted, the phonon modes are displaced in the transformed frames. In the polaron frame, the modes are displaced with $f_\textbf{k}$ depending on the exciton state. In the polariton-polaron frame, the displacement depends on the polariton state $\ket{\pm}$. \label{fig:sketch_system}}
\end{figure}
From the total Hamiltonian, $H=H_{\rm S}+H_{\rm I} + H_{\rm E}$, a master equation can be derived by treating the interaction, $H_{\rm I}$, perturbatively. However, before performing the perturbative expansion, a unitary transformation, $U$, can be applied. If the transformation cannot be factorized into system and environment parts, i.e. $U\neq U_{\rm S}\otimes U_{\rm E}$, it induces a mixing of the system and environmental degrees of freedom. As a result, the transformed Hamiltonian, $\tilde{H} = UHU^\dagger$, needs to be re-partitioned into system, environment and interaction terms.
Even when imposing the Markov approximation in the transformed reference frame it is possible to include non-Markovian effects, because system-environment correlations are built into the transformed basis states. It also follows that the magnitude of the perturbation parameter depends on the transformation and, therefore, different transformations in general lead to different ranges of validity of the ensuing system.

As previously stated, we will consider four master equations in this Article, and these are distinct in the unitary transformation applied to the Hamiltonian prior to derivation of a second-order perturbative Markovian master equation. In this section, we present the underlying unitary transformations and derive the corresponding master equations. The unitary transformations considered are the identity transformation, the standard polaron transformation~\cite{manyparticle,phonon_1,wurger1998strong,duke1965phonon,merrifield1964theory}, a variationally optimized polaron transformation~\cite{Silbey_variational,gomez_2018_variational_master_equation,mccutcheon2011general,nazir2016modelling,Denning_2020_phonon_decoupling} and a new polariton-polaron transformation. An illustration of the exciton-cavity-phonon system together with the effects of the variational polaron transformation and polariton-polaron transformation can be seen in fig. \ref{fig:sketch_system}. Each phonon mode is represented as a harmonic oscillator. The general effect of the transformations is to displace the equilibrium around which the phonon modes oscillate. This is illustrated as a shift in the placement of the harmonic oscillator that represents a phonon mode. As an example, the variational polaron transformation displaces the phonon-modes depending on the state of the exciton, therefore the phonon modes associated with the excited state of the exciton are displaced with factors $f_1,f_2,f_3,...$, while the phonon modes associated with the ground state are not displaced at all. The polariton-polaron transformation describes the system in the polariton frame and displaces the phonon-modes depending on the polariton state, which is depicted accordingly in the figure.

Before deriving the specific master equations by application of these unitary transformations, we briefly outline the general form of second-order perturbative Markovian master equations.

\subsection{Second-order Markovian master equation}
As in section~\ref{sec:model}, we partition the total Hamiltonian into system, environment and interaction parts, $H=H_{\rm S}+H_{\rm E}+H_{\rm I}$. To derive the Markovian master equations used in this paper, a general interaction Hamiltonian which is decomposed into operators working on the system and environment is considered:
\begin{equation}
 \label{eq:Hamiltonian_split}
    H_{\rm I} = \sum_{i}A_i \otimes B_i,
\end{equation}
where $A_i$ works on the system Hilbert space and $B_i$ on the environment Hilbert space. Following e.g. Refs.~\onlinecite{breuer2002theory,nazir2016modelling}, a Markovian second order master equation that is perturbative in $H_{\rm I}$ can be derived:
\begin{align}
\label{eq:general-markov-master-equation}
\begin{split}
    \frac{\mathrm{d}\rho_{\rm S}(t)}{\mathrm{d}t} = &- \frac{i}{\hbar}[H_{\rm S},\rho_{\rm S}(t)] \\ &- \sum_{ij}\int_0^\infty \mathrm{d}\tau\Big\{ C_{ij}(\tau)[A_i,\bar{A}_j(-\tau)\rho_{\rm S}(t)]  \\ &\hspace{1.5cm}+ C_{ji}(-\tau)[A_i,\bar{A}_j(-\tau)\rho_{\rm S}(t)]\Big\}.
\end{split}
\end{align}
Here, we have defined the interaction picture system operators $\bar{A}_i(\tau)=\mathrm{e}^{iH_{\rm S}\tau/\hbar}A_i\mathrm{e}^{-iH_{\rm S}\tau/\hbar}$ and the system part of the density matrix $\rho = \rho_S \otimes \rho_E$ and the environmental correlation functions
\begin{equation}
    C_{ij}(\tau) = \frac{1}{\hbar^2}\mathrm{Tr}_{\rm E} \left[\bar{B}_i(\tau)B_j\rho_{\rm E}(0) \right],
\end{equation}
where $\bar{B}_i(\tau)=e^{iH_{\rm E}\tau/\hbar}B_ie^{-iH_{\rm E}\tau/\hbar}$.
In the derivation of the master equation, a Markovian approximation was imposed by extending the upper limit of the $\tau$-integral in Eq.~\eqref{eq:general-markov-master-equation} from $t$ to infinity~\cite{breuer2002theory}. The environment and system was furthermore assumed to be factorized initially $\rho(0)=\rho_{\rm S}(0)\otimes \rho_{\rm E}(0)$. As a shorthand notation, we shall refer to the terms in the master equation involving the environmental correlation functions as the phononic dissipator, $\mathcal{K}[\rho_{\rm S}]$.

In addition to the phonon-induced effects generated by the Hamiltonian in Section~\ref{sec:model}, we also include a Markovian cavity decay process with rate $\kappa$ as well as pure dephasing resulting from virtual higher-order phonon transitions to energetically higher exciton states \cite{Phonon_dephasing_Reigue,Phonon_dephasing_Tighineanu,Phonon_dephasing_Muljarov} with a rate
\begin{equation}
    \gamma(T) = \frac{\alpha \mu}{v_{\rm c}^4}\int_0^\infty \nu^{10} \mathrm{e}^{-2\nu^2/\nu_{\rm c}}n(\nu)(n(\nu)-1) \ \mathrm{d} \nu,
\end{equation}
where $\mu$ is a material dependent parameter and $n(\nu) = \left(1 - \exp(-\beta\nu) \right)^{-1}$.
Including these effects, the resulting master equation can be written as
\begin{align}
\begin{split}
    \frac{\mathrm{d}\rho_{\rm S}(t)}{\mathrm{d}t} = - \frac{i}{\hbar}[H_{\rm S},\rho_{\rm S}(t)] + \mathcal{K}[\rho] + \kappa \mathcal{D}_{a}[\rho] + 2 \gamma(T) \mathcal{D}_{\sigma^\dagger \sigma}[\rho],
\end{split}
\end{align}
where $\mathcal{D}_A[\rho] = \frac{1}{2}(2A\rho A^\dagger-\rho A^\dagger A - A^\dagger A \rho)$ is the Lindblad dissipator.

We are now in a position to derive the specific master equations resulting from the different unitary transformations.

\subsection{Weak phonon-coupling master equation}
The weak phonon-coupling master equation is the simplest of the four master equations considered, as it does not involve any transformation of the Hamiltonian. The interaction Hamiltonian in eq. \eqref{eq:Hamiltonian_split} thus only has $\{i\} = Z$ and $A_Z = Z = \sigma^\dagger \sigma$ and $B_Z = \sum_\textbf{k}\hbar g_\textbf{k}(b_\textbf{k}^\dagger + b_\textbf{k})$.
The environmental correlation function is found from Ref. \onlinecite{nazir2016modelling}:
\begin{equation}
    C_{ZZ}(\tau) = \int_0^\infty J(\nu)\left[ \coth(\beta \hbar \nu / 2)\cos(\nu \tau) - i \sin(\nu\tau)  \right]. \label{eq:C_ZZ_Weak}
\end{equation}

This outlines the weak phonon-coupling master equation approach. It is worth noting, that the master equation is Markovian, so memory effects due to phonons are lost. Furthermore, the master equation is based on a second order perturbation theory and thus requires the perturbative quantity to be small. The perturbative quantity is the interaction Hamiltonian and it should therefore be small compared to the system and environment Hamiltonian. This means that the coupling rate to phonons should be small compared to the light-matter coupling rate. The weak phonon-coupling master equation is thus perturbative in the phonon coupling, hence its name.

\subsection{Standard and variational polaron master equations}
\label{sec:standard-and-variational-polaron-ME}
In this section, the standard and variational polaron master equations are derived. These transformations are very similar, but differ in that a variational optimization step is performed in the latter, but not the former.
The standard polaron transformation has two main purposes. First of all, the transformation is a way of encoding phonon memory information into the state of the exciton and thus allowing the inclusion of certain non-Markovian effects. This encoding is illustrated in fig. \ref{fig:sketch_system}, where the polaron transformation displaces the phonon-modes depending on the state of the exciton. Secondly, it changes the interaction Hamiltonian from being perturbative in the exciton-phonon coupling to being perturbative in the light-matter coupling \cite{nazir2016modelling}. The standard polaron transformation can thus treat strong exciton-phonon couplings, but instead fail at strong light-matter couplings.

The variational polaron transformation is very similar to the standard polaron transformation, but the transformation leaves the interaction Hamiltonian with terms that can be associated with the weak phonon-coupling master equation and the standard polaron master equation. The variational polaron master equation is therefore a "middleway" between the weak phonon-coupling master equation and the standard polaron master equation. The perturbative parameter of the variational polaron master equation is thus a combination of the light-matter coupling and exciton-phonon coupling, and the transformation is variationally optimized in an attempt to reduce these perturbation terms.

Both polaron transformations are described by the unitary operator $U_{\rm V}=e^{V}$, which transforms the total Hamiltonian as $\tilde{H}_{\rm V}=U_{\rm V} H U_{\rm V}^\dagger$ and is generated by the anti-Hermitian operator \cite{nazir2016modelling}
\begin{align}
\begin{split}
      &V =\sigma^\dagger \sigma \sum_\mathbf{k} \frac{f_\mathbf{k}}{\nu_\mathbf{k}}(b_\mathbf{k}^\dagger - b_\mathbf{k}),
\end{split}
\end{align}
where $\{f_\mathbf{k}\}$ is a set of transformation parameters.
The effect of the transformation and the role of the transformation parameters can be elucidated by expanding $U_\mathrm{v}$ in the form
\begin{align}
\label{eq:variational-polaron-U-expanded}
    U_{\rm V} = \dyad{g} + \dyad{X}\prod_\textbf{k} D_\mathbf{k}(f_\mathbf{k}/\nu_\mathbf{k}),
\end{align}
where $D_{\mathbf{k}}(\alpha_\mathbf{k}) = \exp[\alpha_\mathbf{k}b_\mathbf{k}^\dagger-\alpha_\mathbf{k}^*b_{\mathbf{k}}]$ is the displacement operator of the $\mathbf{k}$th phonon mode.
From this form, it can be seen that the transformation displaces the phonon environment depending on the excitonic state, thus describing polaronic electron--phonon hybridization.

The difference between the standard and variational polaron transformations lies in the choice of the displacement parameters, $f_{\mathbf{k}}$: In the standard polaron transformation, the displacement is fixed at $f_\mathbf{k}=g_\mathbf{k}$, whereby the electron--phonon coupling term vanishes in the transformed Hamiltonian. In the variational polaron transformation, on the other hand, the displacement parameters are determined by minimization of the Bogoliubov upper bound of the free energy.
This variational transformation is similar to the method employed by Silbey et al. studying a spin-boson type model \cite{Silbey_variational,Silbey_1976_comments} which was later adopted by McCutcheon et al. to a coherently driven two-level emitter coupled to a phonon bath to develop a variational master equation \cite{mccutcheon2011general,nazir2016modelling}. The transformation has also been extended to a coherently driven quantized cavity mode coupled to a two-level emitter \cite{gomez_2018_variational_master_equation}. The main physical motivation for this transformation is that it should come as close as possible to diagonalizing the Hamiltonian and thus closely resemble the equilibrium states of the system and reduce the strength of the interaction Hamiltonian \cite{nazir2016modelling}.
To illustrate this, consider the case where the light-matter coupling is vanishing, i.e. $g=0$, in which case the model reduces to the independent boson model~\cite{manyparticle}. Here, the Hamiltonian can be exactly diagonalized through the polaron transformation by setting the transformation variables as $f_\mathbf{k}=g_\mathbf{k}$. If the variational transformation is chosen wisely, then it can approximately diagonalize the Hamiltonian for non-vanishing light-matter couplings $g \neq 0$, only leaving a small interaction term that can be treated perturbatively.
%
%
%
%
To determine the variational parameter $f_\textbf{k}$ such that the transformation results in the smallest interaction term possible, the Bogoliubov upper bound on the free-energy of the system is minimized. To understand why this minimization should lead to a partition of the Hamiltonian with a small interaction term, we consider the Bogoliubov inequality~\cite{BI_1958,BI_2002,Kuzemsky_2015_Variational_principle_of_Bogoliubov} for two Hermitian operators $H$ and $H'$ and follow the arguments of Ref. \onlinecite{Silbey_2008_general_free_energy}:
\begin{equation}
    -\beta^{-1} \ln \mathrm{Tr} \left \{ \mathrm{e}^{-\beta H} \right \}  \leq -\beta^{-1} \ln \mathrm{Tr} \left \{ \mathrm{e}^{-\beta H'} \right \}  + \expval{H - H'}_{H'},
\end{equation}
where $\expval{\cdot}_{H'} = \mathrm{Tr}\left\{ (\cdot) \ \mathrm{e}^{-\beta H'} \right\}/ \mathrm{Tr}\left\{ \mathrm{e}^{-\beta H'} \right\}$ and $\beta = 1/k_b T$. The free energy is unchanged under a general unitary transformation, $UH U^\dagger  = \Tilde{H_0} + \Tilde{H_1}$, and thus choosing $H' = H_0$ gives the inequality:
\begin{equation}
    -\beta^{-1} \ln \mathrm{Tr}\left\{ \mathrm{e}^{-\beta H} \right \} \leq -\beta^{-1} \ln \mathrm{Tr}  \left \{ \mathrm{e}^{-\beta \tilde{H_0}}\right \}+ \langle\tilde{H_1}\rangle
    _{H_0}.
\end{equation}
The inequality only becomes an equality if the unitary transformation diagonalizes the total Hamiltonian, and minimizing the upper bound on the free energy $A_B = -\beta^{-1} \ln \mathrm{Tr}\left \{\mathrm{e}^{-\beta \tilde{H_0}} \right \} + \langle\tilde{H_1}\rangle_{H_0}$ is thus a way of partitioning the Hamiltonian in such a way that the zeroth-order Hamiltonian $H_0$ resembles the diagonalized case as much as possible given the restrictions of the transformation \cite{Silbey_2008_general_free_energy,Kuzemsky_2015_Variational_principle_of_Bogoliubov}.
%
%
%
%

This is the fundamental principle behind the variational polaron approach, where the transformation is applied with the goal of leaving a small interaction term. Applying the transformation to the Hamiltonian leads to the transformed system Hamiltonian (transformed frame denoted by subscript V):
\begin{align}
\begin{split}
\Tilde{H}_{\rm SV} = (\hbar\omega_{eg} + \hbar R)\sigma^\dagger\sigma + \hbar\omega_c a^\dagger a + \hbar g_\mathrm{V}(\sigma^\dagger a + a^\dagger \sigma),
\end{split}
\end{align}
where the variational shift $R = \sum_\mathbf{k} f_\mathbf{k}(f_\mathbf{k}-2g_\mathbf{k})/\nu_\mathbf{k}$ and renormalization $g_\mathrm{V}=\ev{B}g$ has been introduced. $\ev{B} = \ev{B_\pm}$ is the thermal expectation value of the variational displacement operator $B_\pm = \prod_\textbf{k} D_\textbf{k}(\pm \frac{f_\textbf{k}}{\nu_\textbf{k}})$. The frequency shift and coupling-strength renormalisation arise from a rearrangement of terms that assures that the thermal expectation value of the interaction Hamiltonian vanishes, $\ev{H_{\rm IV}} = 0$. 
With this rearrangement, the interaction Hamiltonian in the variational frame is:
\begin{align}
\Tilde{H}_{\rm IV} = XB_X + YB_Y + ZB_Z,
\end{align}
where $X=\hbar g(\sigma^\dagger a + a^\dagger \sigma),\; Y=i\hbar g(\sigma a^\dagger - \sigma^\dagger a),\; Z=\sigma^\dagger\sigma$, and $B_X=(B_++B_--2\ev{B})/2,\; B_Y=i(B_+-B_-)/2,\; B_Z = \sum_\mathbf{k} \hbar(g_\mathbf{k}-f_\mathbf{k})(b_\mathbf{k}^\dagger + b_\mathbf{k})$. $\tilde{H}_{\rm EV}=H_{\rm E}$ is unchanged under the transformation by rearranging terms into $H_{\rm SV}$ and $H_{\rm IV}$.

We here note that taking the limits $f_\textbf{k} \rightarrow 0$ restores the original Hamiltonian which leads to the weak phonon master equation, because the transformation $U_{\rm V}$ reduces to the identity. The other limit of $f_\textbf{k} \rightarrow g_\textbf{k}$ leads to the standard polaron transformation and its corresponding master equation.

Following Refs. \onlinecite{mccutcheon2011general,nazir2016modelling,Denning_2020_phonon_decoupling} and minimizing the upper bound on the free energy leads to the following expressions for the variational factors $f_\textbf{k}$ and the resulting frequency shift and light-matter renormalization:
\begin{align}
\frac{f_\mathbf{k}}{g_\mathbf{k}} &= \frac{\qty[1 - \frac{\delta_\mathrm{v}}{\eta_\mathrm{v}}\tanh(\beta\hbar\eta_\mathrm{v}/2)]}{1 - \frac{\delta_\mathrm{v}}{\eta_\mathrm{v}}\tanh(\beta\hbar\eta_\mathrm{v}/2)\qty[1 - \frac{2g_\mathrm{V}^2}{\nu_\mathbf{k}\delta_\mathrm{v}}\coth(\beta\hbar\nu_\mathbf{k}/2)]}\label{eq:var_param},\\
R &= \int_0^\infty\dd{\nu} \frac{J(\nu)}{\nu}F(\nu)[F(\nu)-2],
\label{eq:R_V} \\
\ev{B} &= \exp[-\frac{1}{2}\int_0^\infty \dd{\nu}\frac{J(\nu)F^2(\nu)}{\nu^2}\coth(\beta\hbar\nu/2)],
\label{eq:R_B}
\end{align}
where the sums have been converted into integrals using the spectral density function $J(\nu)$ and the dimensionless variational function $F(\nu_\textbf{k}) = \frac{f_\textbf{k}}{g_\textbf{k}}$ has been defined as well as the quantities $\eta_{\mathrm{V}}=\sqrt{4g_\mathrm{V}^2 + \delta_\mathrm{V}^2},\; \delta_\mathrm{V} = \omega_{eg}+R - \omega_c:=\Delta + R$. The variational equation, Eq.~\eqref{eq:var_param}, thus expresses an implicit relation for $F(\nu)$, which needs to be solved self-consistently. This self-consistent solution is obtained numerically.

Using the definition in Eq. \eqref{eq:Hamiltonian_split}, the interaction Hamiltonian now has $i =X,Y,Z$
with $A_i = X,Y,Z$ and $B_i = B_X,B_Y,B_Z$. The phonon correlation functions $C_{ij}(\tau)$ are all given from Ref.~\onlinecite{nazir2016modelling} as
\begin{align}
\label{eq:C_ij-variational}
  \begin{split}
    C_{XX}(\tau) &= \frac{\ev{B}^2}{2}(e^{\phi(\tau)} + e^{-\phi(\tau)} - 2), \\
    C_{YY}(\tau) &= \frac{\ev{B}^2}{2}(e^{\phi(\tau)} - e^{-\phi(\tau)}), \\
    C_{ZZ}(\tau) &= \int_0^\infty\dd{\nu}J(\nu)[1-F(\nu)]^2 \cdot
    \\&[\coth(\beta\hbar\nu/2)\cos(\nu\tau) -i\sin(\nu\tau)], \\
    C_{YZ}(\tau) &= -\ev{B}\int_0^\infty \dd{\nu} J(\nu)\nu^{-1}F(\nu)[1-F(\nu)] \cdot \\
    &[i\cos(\nu\tau) + \coth(\beta\nu/2)\sin(\nu\tau)], \\
    C_{ZY}(\tau) &= -C_{YZ}(\tau),
  \end{split}
\end{align}
where $\phi(\tau)=\int_0^\infty\dd{\nu}J(\nu)\nu^{-2}F^2(\nu)[\coth(\beta\hbar\nu/2)\cos(\nu\tau) -i\sin(\nu\tau)]$ and $C_{XY},\;C_{YX},\;C_{XZ}$ and $C_{ZX}$ are zero.

We note that the standard polaron master equation can directly be obtained from the variational master equation by setting $F(\nu) = 1$ rather than determining $F(\nu)$ through Eq.~\eqref{eq:var_param}. This leaves only $C_{XX}$ and $C_{YY}$ as non-zero correlation functions. Similarly, the weak phonon-coupling master equation can be recovered by setting $F(\nu) = 0$, in which case only $C_{ZZ}$ is nonzero.

\subsection{Polariton-Polaron transformation}
In the absence of interactions with the environment, the eigenstates of the Jaynes-Cummings Hamiltonian are known as the dressed states or upper and lower polariton states \cite{jaynes1963comparison}. In the strong-coupling regime, which can be roughly estimated as for $4g > \kappa$, these polariton states dominate the optical response of the system. Describing the system in a basis of the polariton states is therefore natural when the system is in the strong-coupling regime. In the previous section, the variational polaron transformation was introduced. This approach is based on an assumption of how the equilibrium states approximately look for non-vanishing light-matter couplings. A central feature of the polaronic transformations is that they are diagonal in the uncoupled exciton basis, $\{\ket{g}, \ket{X}\}$, as is explicitly seen in Eq.~\eqref{eq:variational-polaron-U-expanded}. However, since the eigenstates of the system in the absence of phonon interactions are the polariton states, it seems natural to introduce a transformation that is diagonal in the dressed exciton-cavity states rather than the bare exciton states.

In this section, we introduce such a transformation and its associated master equation approach. Here, the system is described in the basis of the polariton states and a transformation that dresses these polariton states with polarons is performed. The approach is inspired by recent studies showing optical signatures of dressed polaritons \cite{Denning_2020_phonon_decoupling} or polaron-polaritons \cite{Wu_2016_polarons_meet_polaritons}. The transformation is similar to that of Ref. \onlinecite{Wu_2016_polarons_meet_polaritons}, where the cavity modes are explicitly dressed by a generalized Merrifield transformation. The transformation we propose is, however, more intuitive since it is described in the polariton basis and the dressing or formation of polariton-polarons can be explicitly seen. Furthermore, the necessary groundwork for a master equation description is also established.

We start by writing our Hamiltonian in the basis of the polaritons. The polariton states are given as \cite{jaynes1963comparison}:
\begin{align}
\begin{split}
    &\ket{+} = C_+ \ket{X,0} + C_- \ket{g,1} \\
    &\ket{-} = C_- \ket{X,0} - C_+ \ket{g,1}
\end{split}
\end{align}
where $C_\pm = \frac{1}{\sqrt{2}}\left( 1 \pm \frac{\Delta}{(\Delta^2 + 4g^2)^{1/2}}  \right)^{1/2}$ and the exciton-cavity basis states are defined as $\ket{\alpha,n}=\ket{\alpha}\otimes\ket{n}$, where $\alpha=g,X$ and $n=0,1$ are the exciton state and cavity photon number, respectively. For the sake of simplicity, it will be assumed that there is no detuning $\Delta = 0$ and thus $C_+=C_-=\frac{1}{\sqrt{2}}$. The following approach can straightforwardly be extended to the non-zero detuning case, but this will require a calculation of a new spectral density-like function and the expressions are generally cumbersome.

The polaritons are the eigenstates of the Jaynes-Cummings Hamiltonian and the system Hamiltonian is therefore diagonal in the polariton states with their respective eigenvalues $E_\pm = (\hbar\omega_{eg}+ \hbar\omega_{c})/2 \pm \hbar g$. The interaction Hamiltonian, however, contains the term $\sigma^\dagger \sigma = \ket{X}\bra{X}$ which needs to be written in terms of the dressed states. We assume that there is maximally one excitation in the system and thus $\ket{X}\bra{X} \otimes \ket{1}\bra{1}$ is not a possible state which reduces $\sigma^\dagger \sigma$ to: $\sigma^\dagger \sigma = \frac{1}{2}( p^\dagger p + m^\dagger m - p^\dagger m - m^\dagger p)$ where the notation $p  = \ket{g,0}\bra{+}$ and $m = \ket{g,0}\bra{-}$ has been introduced. The total Hamiltonian can thus be expanded in the dressed state basis as
\begin{align}
\begin{split}
    H = & E_+ p^\dagger p + E_- m^\dagger m + \sum_\mathbf{k} \hbar \nu_\mathbf{k} b_\mathbf{k}^\dagger b \\ + &\frac{1}{2}\left( p^\dagger p + m^\dagger m - p^\dagger m + m^\dagger p \right) \sum_\mathbf{k}\hbar g_\mathbf{k}(b_\mathbf{k}^\dagger + b_\mathbf{k}). \label{eq:polariton_Hamiltonian}
\end{split}
\end{align}
At this point, we define a unitary transformation of the form
\begin{align}
     W = &\ket{g,0}\bra{g,0} + m^\dagger m \exp[\sum_\mathbf{k} \frac{f_\mathbf{k}^-}{\nu_\mathbf{k}} (b^\dagger-b)]\\ + &p^\dagger p \exp[\sum_\mathbf{k} \frac{f_\mathbf{k}^+}{\nu_\mathbf{k}} (b^\dagger-b)].
     \label{eq:polariton_transformation}
\end{align}
This transformation generates a polaronic phonon-displacement that is diagonal in the polariton basis, and we thus denote it as the \emph{polariton polaron transformation}. In analogy with the standard polaron transformation, the coefficients $f_\mathbf{k}^\pm$ are determined such that the phonon interaction terms that are diagonal in the polariton basis are eliminated. In appendix \ref{app:Polariton_polaron}, the transformation is applied and for the resonant case ($\Delta=0$), we find this requirement to be $f_\mathbf{k}^-=f_\mathbf{k}^+=g_\mathbf{k}/2$, yielding the transformed Hamiltonian
\begin{align}
\begin{split}
    &\Tilde{H}_{\rm SW} = p^\dagger p \left( E_+ + \frac{\hbar \Delta_p}{4} \right) + m^\dagger m \left(E_- + \frac{\hbar \Delta_p}{4}\right)  \\
    & - \frac{\hbar\Delta_p}{2} \left( p^\dagger m  + m^\dagger p \right), \\ &\Tilde{H}_{\rm IW} = \sum_\mathbf{k} - \frac{\hbar g_\mathbf{k}}{2} (b_\mathbf{k}^\dagger + b_\mathbf{k}) \left[ p^\dagger m + m^\dagger p \right].
\end{split}
\end{align}
The environment Hamiltonian, $\tilde{H}_{\rm EW}=H_{\rm E}$ is again unchanged by rearranging terms. Using the definition in eq. \eqref{eq:Hamiltonian_split}, the interaction Hamiltonian has one term with $A_i = -\frac{1}{2}(p^\dagger m + m^\dagger p)$ and $B_i = B_Z = \sum_\textbf{k} \hbar g_\textbf{k} (b_\mathbf{k}^\dagger + b_\mathbf{k})$. We note that phonon part $B_Z$ is identical to the one appearing in the weak coupling master equation and the phonon correlation function is already known from Eq. \eqref{eq:C_ZZ_Weak}. This concludes the derivation of the master equations. Very importantly, however, three of the four master equations describe the system evolution in a transformed frame. This has crucial implications on how the emission spectrum should be calculated and this will be elaborated upon in the next section.

\section{Emission spectrum}
There are generally two approaches for calculating the emission spectrum of the cavity from one of the perturbative master equations. Both involve calculating a two-time correlation function from the quantum regression theorem \cite{carmichaael}:
\begin{equation}
    \langle O^\dagger (t+\tau) O(t) \rangle = \mathrm{Tr} \left\{ O^\dagger \mathrm{e}^{\mathcal{L}\tau} O \mathrm{e}^{\mathcal{L}t} \rho(0) \right\}, \label{eq:QRT}
\end{equation}
where $O$ is an arbitrary operator and $\mathcal{L}$ is a Liouvillian superoperator defined so that the time evolution of the density matrix is given as $\frac{\mathrm{d}\rho}{\mathrm{d}t} = \mathcal{L}[\rho]$. $\rho(0)$ is set to $\rho(0) = \ket{X}\bra{X}$ which mimics the moment after the emitter has been excited. The difference between the two methods for calculating the emission spectrum is the operator, $O$, that is used for calculation of the two-time average. Specifically, the cavity emission spectrum can either be directly calculated from the Fourier transform of the cavity correlation function by setting $O=a$. Alternatively, the spectrum can be calculated from the dipole spectrum ($O=\sigma$) via a Green's function. As we shall see, in the limit of weak light-matter coupling $4g<\kappa$, this dipole-based calculation is preferable for the polaron-type master equations described in Sec.~\ref{sec:standard-and-variational-polaron-ME} because it allows to capture non-Markovian effects manifested in the spectral phonon sideband~\cite{Hughes_2015_dipole_emission}.

We note that the quantum regression theorem only holds when the time evolution is Markovian~\cite{carmichaael,breuer2002theory}, but since the two-time average in Eq.~\eqref{eq:QRT} is evaluated in a transformed reference frame, this does not imply that Markovian evolution in the original (laboratory) reference frame is needed. As will be demonstrated in the following, pronounced non-Markovian effects arising from the delayed phonon response can still be captured. Other studies have examined the validity of the quantum regression theorem in the laboratory frame and to some extent under the standard polaron transformation~\cite{cosacchi2021accuracy}. We find that the Markov approximation and thus the applicability of the quantum regression theorem is closely connected to the suitability of the applied basis transformation and the perturbation strength in the transformed reference frame.

\subsection{Dipole spectrum method}
\label{sec:dipole-spectrum-method}
The cavity emission spectrum can be related to the dipole correlation function as \cite{Hughes_2015_dipole_emission,simultanoeusefficiency}:
\begin{equation}
    S(\omega) =  \mathcal{G}(\omega) \int_{-\infty}^\infty \mathrm{d}t  \int_{-\infty}^\infty \mathrm{d}\tau \ \mathrm{e}^{-i\omega\tau} \langle \sigma^\dagger(t+\tau)\sigma(t)\rangle.
    \label{eq:emissionspec_general}
\end{equation}
where $\mathcal{G}(\omega)$ is the optical Green's function connecting the dipole emission spectrum to the cavity emission spectrum, which in the present case of a single-mode cavity is given by~\cite{simultanoeusefficiency} $G(\omega) = \frac{4g^2}{\kappa} \frac{(\kappa/2)^2}{\omega^2 + (\kappa/2)^2} \label{eq:green}$. In the evaluation of the two-time correlation function through a master equation formulated in a transformed reference frame, it is necessary to transform the correlation function back to the 'lab' or 'original' reference frame~\cite{Hughes_2015_dipole_emission}. For the standard and variational polaron transformations this gives~\cite{phononeffects_emil}
\begin{equation}
    \expval{\sigma^\dagger(t+\tau)\sigma(t)} = \expval{\sigma^\dagger(t+\tau) B_-(t+\tau) B_+(t) \mathrm{e}^{Q(t)} \sigma(t)}_{\rm V}, \label{eq:polaron_frame_correlation}
\end{equation}
where the subscript V denotes evaluation of the expectation value in the polaron frame.

Assuming that the phonon bath is in thermal equilibrium in the transformed reference frame, the two-time average can be split into a thermal phonon part and an emitter-cavity part \cite{phononeffects_emil,Iles_Smith_2017_Limits_to_coherent}:
\begin{equation}
\label{eq:exp-val-dipole-lab-frame-polaron}
    \expval{\sigma^\dagger(t+\tau)  \sigma(t)} \approx  \ev{B}^2 \mathrm{e}^{\phi(\tau)} \expval{\sigma^\dagger(t+\tau) \sigma(t)}_{\rm V}.
\end{equation}
The two-time average $\expval{\sigma^\dagger(t+\tau) \sigma(t)}_{\rm V}$ can be evaluated directly from the standard or variational polaron master equation using Eq.~\eqref{eq:QRT}. This relation shows how the standard and variational polaron approaches allows to resolve phononic memory effects in the time evolution of the system, when the Markov approximation is imposed in the transformed reference frame. When the dipole operator is transformed back to the original reference frame, a displacement of the phonon bath in the correlation function is included thus carrying information about the phonon bath. It is the phonon correlation function in Eq.~\eqref{eq:exp-val-dipole-lab-frame-polaron} that gives rise to the non-Markovian phonon sideband in the emission spectrum~\cite{Hughes_2015_dipole_emission}.

For the polariton-polaron transformation, the dipole operator transforms as $\sigma  \rightarrow C_+ p \mathrm{e}^{-Q_+} -  C_- m \mathrm{e}^{-Q_-}$ with $Q_\pm = \sum_\textbf{k} \frac{f_{\textbf{k}}^\pm}{\nu_\textbf{k}}(b_\textbf{k}^\dagger  - b_\textbf{k} )$. The two-time correlation function is therefore written as
\begin{align}
\begin{split}
    &\expval{\sigma^\dagger(t)\sigma(t')} = \left\langle \left( C_+ \mathrm{e}^{Q_+(t)} p^\dagger(t) - C_- \mathrm{e}^{Q_-(t)} m^\dagger(t) \right) \right. \\
    & \left. \left(C_+ \mathrm{e}^{-Q_+(t')} p^\dagger(t') - C_- \mathrm{e}^{-Q_-(t')} m^\dagger(t') \right)\right\rangle_{\rm W}, \label{eq:polariton_frame_correlation}
\end{split}
\end{align}
where the subscript W signifies that the expectation value is evaluated in the polariton-polaron reference frame, i.e. under the transformation $W$. As for the standard and variational polaron transformations, we assume that the phonon environment remains close to its thermal state in the transformed reference frame, thereby allowing to factorize the correlation function. With zero detuning ($\Delta=0$) we have $C_+=C_-$ and $f_\mathbf{k}^+ = f_\mathbf{k}^-$, which leads to:
\begin{equation}
    \expval{\sigma^\dagger(t+\tau)\sigma(t)} \approx \tilde{B}^{1/2}\mathrm{e}^{\tilde{\phi}(\tau)/4} \expval{\sigma^\dagger(t+\tau)\sigma(t)}_{\rm W},
\end{equation}
where $\tilde{B}^{1/2}$ and $\tilde{\phi}(\tau)$ are defined as in eq. \eqref{eq:R_B} and below eq.~\eqref{eq:C_ij-variational}, with $F(\nu)=1$.

\subsection{Cavity spectrum method}
\label{sec:cavity-spectrum-method}
The other and more straightforward method to calculate the emission spectrum is to use the cavity operator \cite{Kaer_2014_cav_emission,Hornecker_2017_cav_emission}:
\begin{equation}
    S(\omega) = \kappa \int_{-\infty}^\infty \int_{-\infty}^\infty \langle a^\dagger(t + \tau) a(t) \rangle \mathrm{e}^{- i \omega \tau} \mathrm{d}t \mathrm{d}\tau.
\end{equation}
The cavity operator is unchanged under the standard and variational polaron transformations and therefore no further work is required in this case, i.e. $\langle a^\dagger(t + \tau) a(t) \rangle=\langle a^\dagger(t + \tau) a(t) \rangle_{\rm V}$. This also means that no phonon-sideband effects can be captured in this approach. As we shall see, the dipole-spectrum approach breaks down in the strong light-matter coupling regime, whereby the consideration of the cavity-spectrum becomes the only feasible approach.

However, under the polariton-polaron transformation, the cavity operator transforms as $a \rightarrow C_- p \mathrm{e}^{-Q_+} +  C_+ m \mathrm{e}^{-Q_-}$. Thus, writing out the correlation function in the original reference frame in terms of the transformed quantities and factorizing the phononic part in complete analogy with the procedure in Sec.~\ref{sec:dipole-spectrum-method} yields
\begin{equation}
    \expval{a^\dagger(t+\tau)a(t)} \approx \tilde{B}^{1/2}\mathrm{e}^{\tilde{\phi}(\tau)/4} \expval{a^\dagger(t+\tau)a(t)}_{\rm W}.
\end{equation}
When the cavity operator method is used, the polariton-polaron method is therefore the only master equation able to incorporate phonon memory effects. As we shall see, this means that phonon sideband effects can be well-described even in the regime of strong light-matter coupling.

With the necessary theoretical groundwork established, we are now ready to benchmark the four master equations against tensor-network reference calculations.

\section{Benchmark of emission spectrum}
\begin{figure*}
    \centering
    \includegraphics[width = \textwidth]{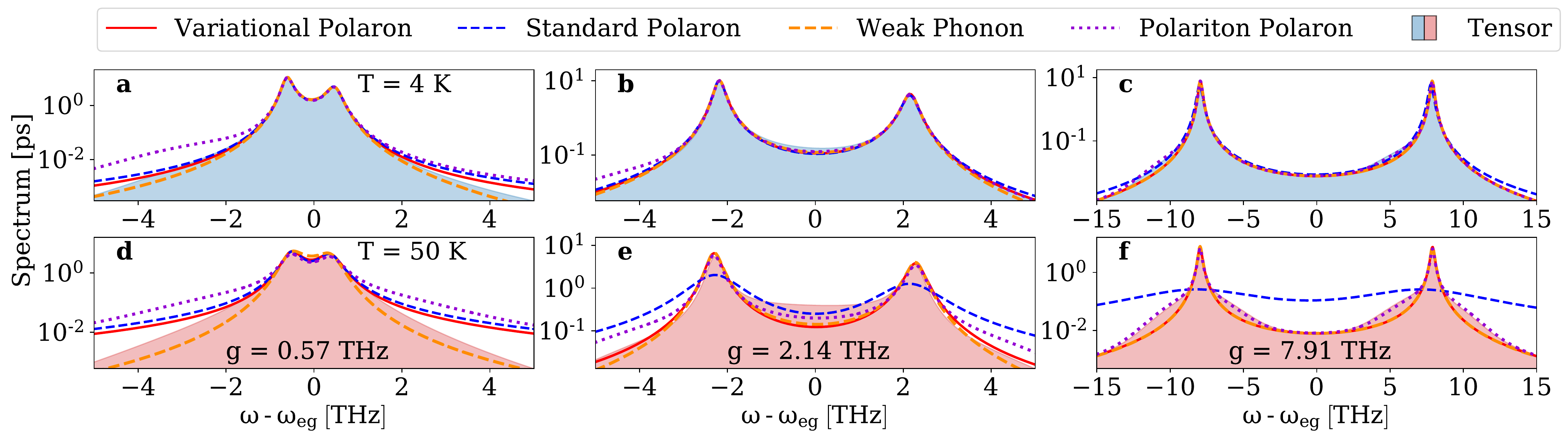}
    \caption{The cavity emission spectrum, $S(\omega)$, calculated from the four master equations resulting from different transformations (lines) and compared to numerically accurate tensor network calculations (shaded graph). The upper panels show the calculations for $T=4 \ \mathrm{K}$ and the lower for $T=50 \ \mathrm{K}$. The cavity decay rate is $\kappa = 0.5 \ \mathrm{THz}$ and the light-matter coupling is varied from $0.57$ to $7.91 \ \mathrm{THz}$.}
    \label{fig:cav_emission}
\end{figure*}

In this section, we benchmark calculations of the emission spectrum using the four master equations introduced against the tensor-network reference calculation. An estimate of the accuracy of the tensor network calculations together with an elaboration of the approach is given in Appendix~\ref{app:Tensor_network}. In most cases, the accuracy of the tensor network is on the order of 1\% or better, which is accurate enough to determine the validity of the master equations.

As a concrete example, we consider a quantum dot with a confinement length of 3 nm embedded in GaAs as the emitter which leads to the following phonon parameters \cite{phononeffects_emil}: Phonon coupling constant $\alpha = 0.0251 \ \mathrm{THz}$, phonon cutoff frequency $\nu_c = 2.23 \ \mathrm{THz}$, and pure dephasing constant $\mu = 0.02284 \ \mathrm{ps}^4$. These parameters will be used throughout the paper.

\subsection{Cavity emission spectrum}
\label{sec:benchmark-cavity-spec}
We start by considering the the cavity operator approach for calculating the emission spectra described in Sec.~\ref{sec:cavity-spectrum-method}, in the strong light-matter coupling regime. We set $\kappa = 0.5 \ \mathrm{THz}$ and vary the light-matter coupling $g$ from $0.57$ to $10 \ \mathrm{THz}$. This range of light-matter couplings represent possible values obtainable through dielectric bow-tie cavities with deep sub-wavelength confinement \cite{Denning_2020_phonon_decoupling,self_similar_Choi_2017}. The range of light-matter couplings also investigate the transition from a configuration where the phonon environment has the fastest timescale to a configuration where the light-coupling rate exceeds the frequency of the phonon environment. This type of configuration is relevant in a number of other physically realizable experimental platforms such as Transition metal dichalcogenides \cite{kleemann2017strong,geisler2019single,qin2020revealing}, Single methylene blue molecules \cite{chikkaraddy2016single} and Nitrogen-vacancy centers \cite{NV_center1,NV_center2}.

The resulting emission spectra can be seen in Fig. \ref{fig:cav_emission} for cryogenic temperatures $T=4\ \mathrm{K}$, and high temperatures $T=50 \ \mathrm{K}$. In the former, the asymmetries due to phonons are pronounced and, in the latter, the thermal energy is high enough to excite a substantial phonon population which leads to stronger but spectrally symmetric phonon effects. The doubled peaked structure of the emission spectra show that the system is in the strong-coupled regime, where polaritons form. For $T=4 \ \mathrm{K}$, the asymmetric phonon effects manifest themselves in uneven heights of the polariton peaks. The left peak is higher than the right because the phonon emission process $\ket{+} \rightarrow{} \ket{-}$ is dominating over the absorption process $\ket{-} \rightarrow{} \ket{+}$ due to the low temperature and, therefore, small population of the phonon modes \cite{Denning_2020_phonon_decoupling}.

For $T=50 \ \mathrm{K}$, the formation of polariton-polarons is pronounced due to the stronger exciton-phonon interaction associated with higher temperatures. This is seen in Fig. \ref{fig:cav_emission}f. The hybridization leading to the double peaked structure constitutes the polaritons, while the forming of polarons is associated with the sideband structure observed around each of the peaks \cite{Wigger_2020_polaron_sideband}. Only the polariton-polaron master equation is able to capture this sideband structure and the violet-dotted line in fig. \ref{fig:cav_emission}f therefore also serve to highlight the appearance of this phonon sideband. The motivation behind the polariton-polaron transformation is now clear. By dressing each of the polariton states with phonons, we can capture the highly non-Markovian phonon-sidebands. This is similar to the polaron transformation, but the basis of the transformation is different, which successfully captures the phonon sidebands of the polariton peaks. The polariton-polaron master equation and the exact tensor network calculation together show that the emission spectrum of a strongly coupled cavity can be explained by the forming of polariton-polarons.

The master equations have tails on their emission spectra that deviate from the tensor-network. We emphasize that the tails are not originating from inaccuracies in the numerical solution of the master equations, but arise due to the perturbative terms in the Liouvillian.

\begin{figure}
    \centering
    \includegraphics[width = 0.99 \linewidth]{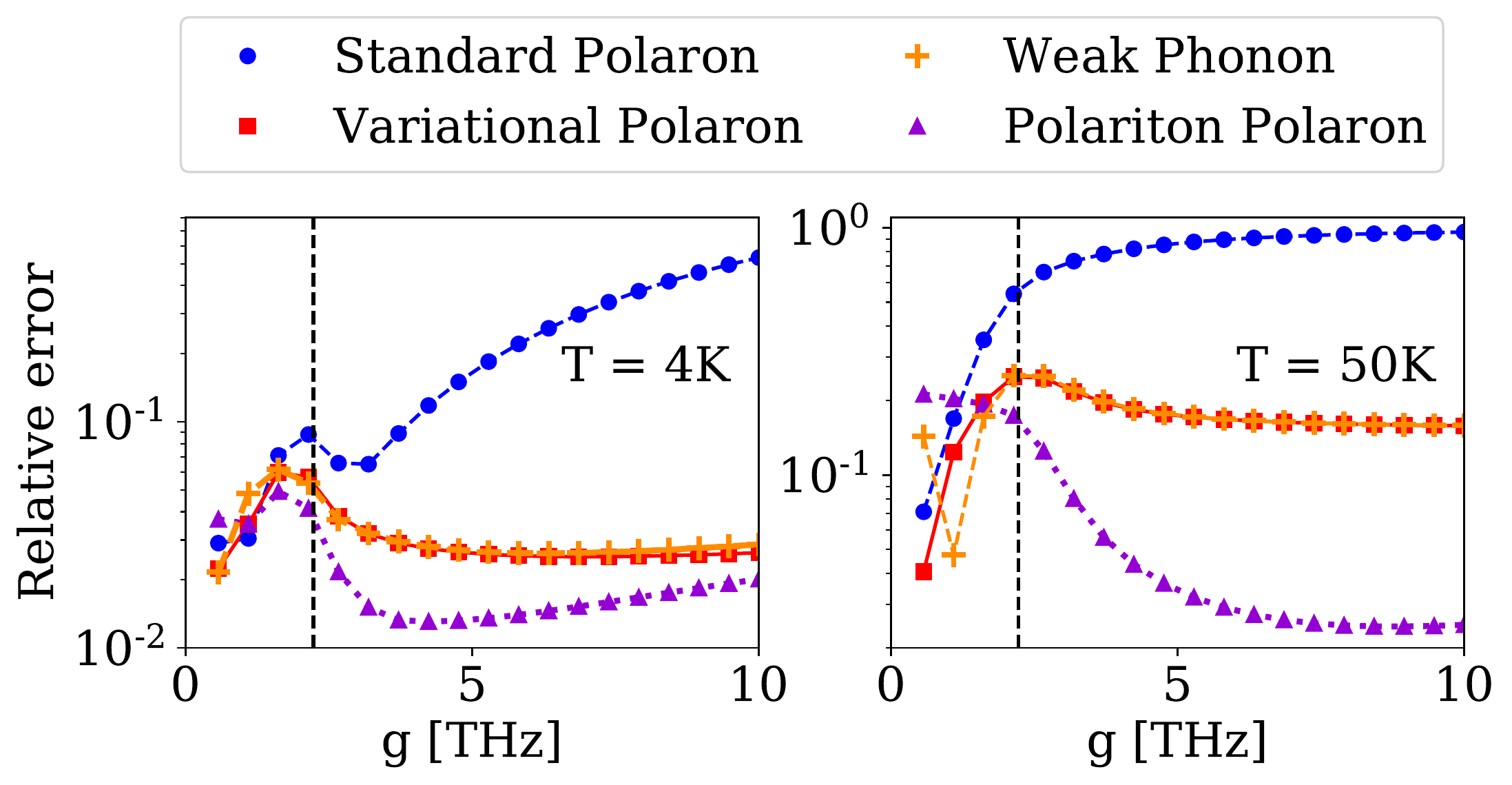}
    \caption{The relative mean square error for the variational, polaron and weak phonon methods at $T=4 \ \mathrm{K}$ ($T=50 \ \mathrm{K}$) for left (right) panel. The dotted vertical lines indicate the phonon cutoff frequency $g=\nu_c=2.23 \ \mathrm{THz}$.}
    \label{fig:error_T4}
\end{figure}
To further assess the accuracy of the different master equations, we compute the root mean square relative error with respect to the tensor network calculation:
\begin{align}
    \Delta S = \left(\frac{\int |S_{\rm tn}(\omega)-S(\omega)|^2 \ \mathrm{d}\omega}{\int |S_{\rm tn}(\omega)|^2 \ \mathrm{d}\omega}\right)^{1/2},
    \label{eq:rel_error}
\end{align}
where $S_{\rm tn}(\omega)$ is the tensor network spectrum and $S(\omega)$ the relevant master equation spectrum. The spectral integration area is from $-62.8 \ \mathrm{THz}$ to $62.8 \ \mathrm{THz}$ and thus all important features are included. The relative error can be seen in Fig. \ref{fig:error_T4}. The variational polaron approach is seen to have a lower or approximately equal relative error compared with the standard polaron and weak coupling master equation at all light-matter couplings. The minimization of the free energy has thus led to an improved perturbation theory within the restrictions given by the variational polaron transformation. From Fig. \ref{fig:error_T4} it is also clear that the polariton-polaron master equation performs better than all the other approaches when the light-matter coupling exceeds the phonon cutoff frequency (marked with dashed lines). This is especially true when the temperature is high, and the polariton-sidebands are pronounced.

\begin{figure}
    \centering
    \includegraphics[width = \linewidth]{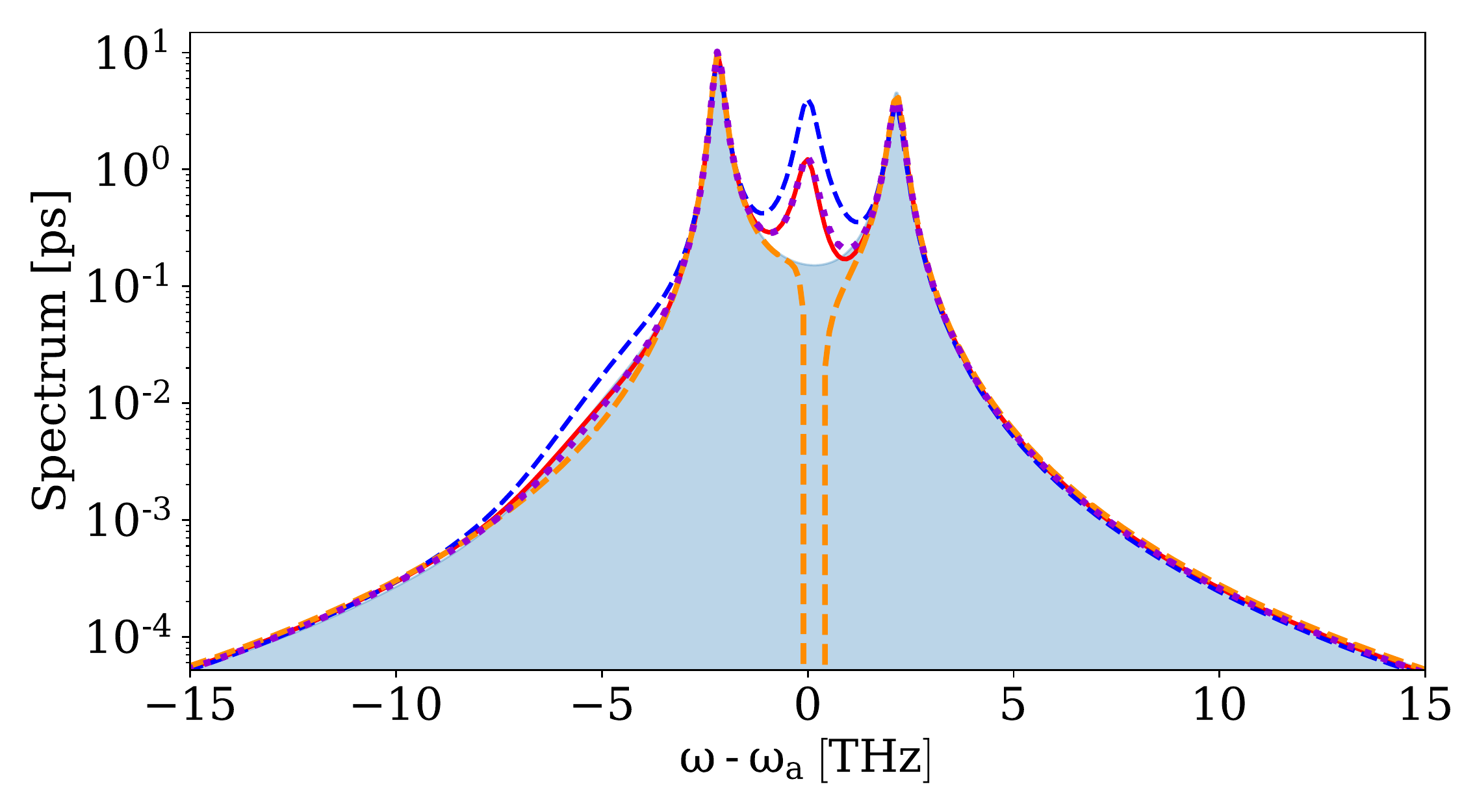}
    \caption{The emission spectrum of the four master equations calculated with the dipole operator method in the strong-coupling regime with $\kappa = 0.5 \ \mathrm{THz}$ and $g=2.14 \ \mathrm{THz}$. It is clear that all 4 master equations produce an incorrect feature close to zero frequency detuning that is not predicted by the tensor-network. Linestyles are the same as in fig. \ref{fig:cav_emission} and \ref{fig:p4_emission}.}
    \label{fig:unphysical}
\end{figure}

\begin{figure*}[t]
    \centering
    \includegraphics[width=\textwidth]{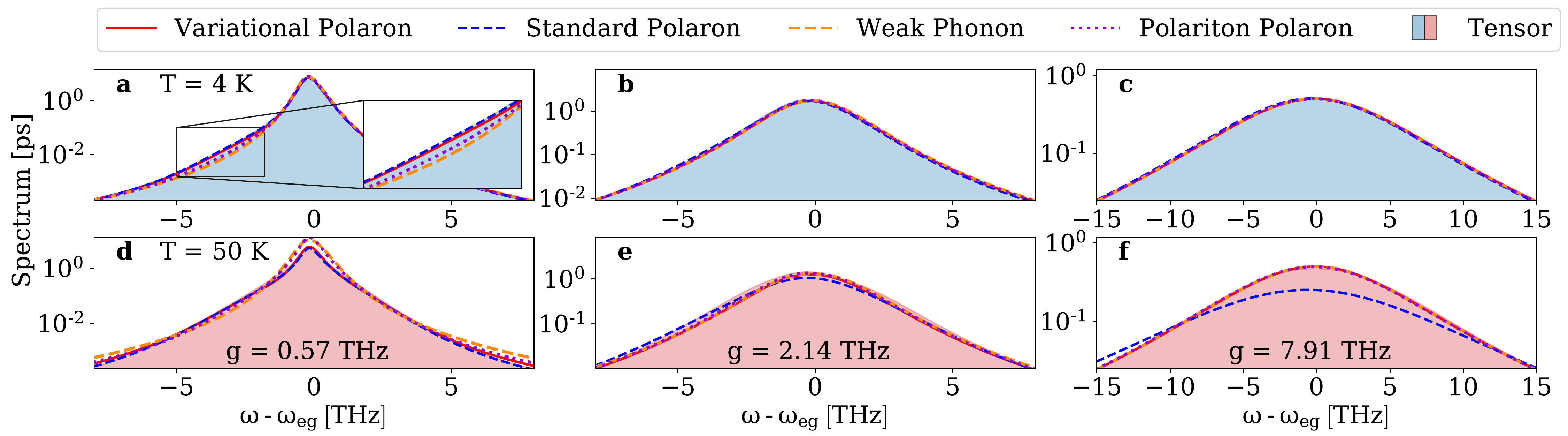}
    \caption{The emission spectrum calculated via the dipole operator method from the four master equations resulting from different transformations (lines) and compared to numerically accurate tensor network calculations (shaded graph). The system is kept in the Purcell regime by setting $\kappa = 4g$. The light-matter coupling rate is varied from $0.57 \ \mathrm{THz}$ to $7.91 \ \mathrm{THz}$. The upper panels show spectra at $T=4 \ \mathrm{K}$ and lower panels show $T=50 \ \mathrm{K}$.}
    \label{fig:p4_emission}
\end{figure*}

Besides introducing broad spectral features and asymmetries into the emission spectra, phonons also shift the emitter frequency and thus change the resonance condition of the system. This is seen explicitly in the variational and standard polaron transformations with the introduction of the shift $R$. This quantity does, however, not fully describe the spectral shift due to phonons. As shown in appendix \ref{app:split_and_shift} a second-order phonon-induced Lamb-shift plays an important role in estimating the spectral shift due to phonons. Another effect of phonons is an effective change in the light-matter coupling $g$, shown explicitly by $B$ in the variational and standard polaron transformation. This renormalization of the light-matter coupling is again best captured by including second-order effects introduced by the phonon-dissipator.

%

\subsection{Dipole emission spectrum}
In Sec.~\ref{sec:benchmark-cavity-spec}, the cavity operator was used for calculating the emission spectrum and the polariton-polaron approach was seen to be superior when the light-matter coupling exceeded the phonon cutoff frequency. Importantly, the polariton-polaron transformation modifies the cavity operator itself, which introduces non-Markovian phonon effects in the sidebands. The variational and standard polaron transformations, in contrast, do not change the cavity operator, and non-Markovian sideband effects cannot enter the cavity operator approach. Using the dipole operator is not an option when considering a strongly coupled cavity, because the method is well known to produce a spurious and non-physical peak at the cavity frequency in this regime  \cite{Hughes_2015_dipole_emission}. This is also illustrated in Fig. \ref{fig:unphysical}, where the cavity decay rate is $\kappa = 0.5 \ \mathrm{THz}$ and $g = 2.14 \ \mathrm{THz}$. The dipole operator method is therefore not suitable for studying the strong-coupling regime and instead the cavity operator method must be used. This limits the applicability of the standard and variational polaron approaches, when considering a strongly coupled cavity.

The dipole operator method is, however, useful for analyzing the Purcell regime where the non-physical peak does not appear. To assess the accuracy of the master equation approaches in the Purcell regime, we pin the cavity decay rate relative to the coupling strength so that $\kappa = 4g$ and vary the light-matter coupling $g$ from $0.57 \ \mathrm{THz}$ to $10 \ \mathrm{THz}$ as in Sec.~\ref{sec:benchmark-cavity-spec}. The emission spectra can be seen in Fig. \ref{fig:p4_emission}. Here, an asymmetric phonon-sideband is visible in Fig. \ref{fig:p4_emission}a.
The phonon sideband results from polaronic phonon-dressing of the excitonic transition, which leads to phonon-assisted relaxation of the exciton. As discussed in Sec.~\ref{sec:dipole-spectrum-method}, it is possible to capture such non-Markovian effects in the variational and standard polaron master equations, because the inverse transformation from the polaron reference frame is accompanied by a phonon correlation function. This is illustrated in Fig.~\ref{fig:p4_emission}a, where the sideband contribution to the spectrum is accurately captured by the standard and variational polaron master equations. The sideband is, on the other hand, not captured by the weak phonon-coupling and polariton-polaron master equations. The weak phonon-coupling master equation had no transformation involved and these non-Markovian effects are therefore lost. The polariton-polaron master equation does employ a transformation, but this transformation dresses the polariton states with phonons and these are not the relevant eigenstates in the Purcell regime.

When the light-matter coupling strength is increased, the phonons are dynamically decoupled \cite{Denning_2020_phonon_decoupling}, because the exciton relaxation rate due to radiative transitions is strongly increased due to the Purcell-effect, while the phonon scattering rate is largely unaffected. This means that the phonon side-band becomes less important, as seen in Fig. \ref{fig:p4_emission}b-c. At $T=50 \ \mathrm{K}$ the phonon-sideband is no longer asymmetric which is seen in Fig. \ref{fig:p4_emission}d.

\begin{figure}
    \centering
    \includegraphics[width = \linewidth]{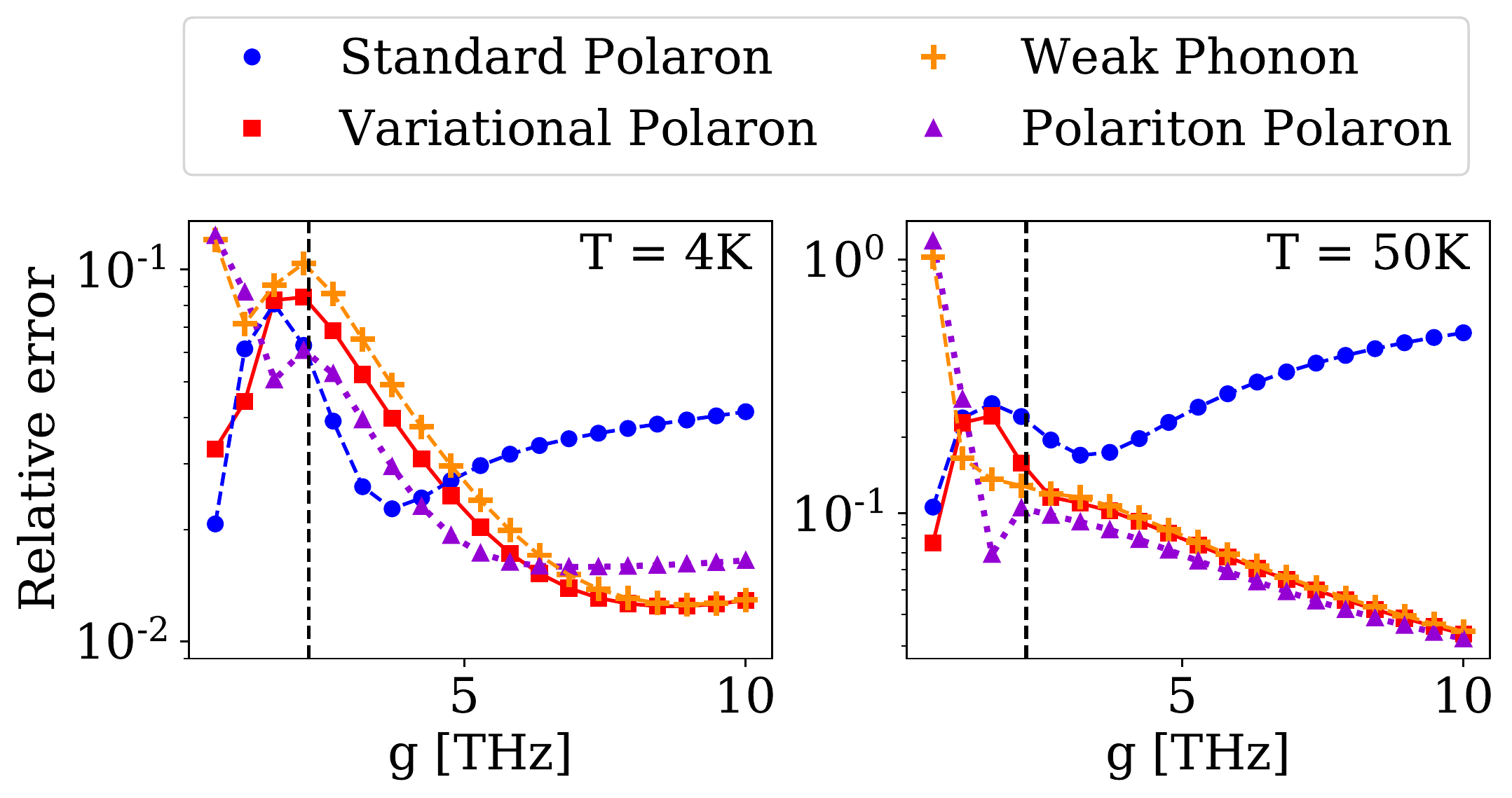}
    \caption{Relative root mean square error for the same parameters as in Fig. \ref{fig:p4_emission}, calculated as in eq. \eqref{eq:rel_error}.}
    \label{fig:error_dipole_T4}
\end{figure}

The phonon decoupling is also illustrated in the relative error plot shown in fig. \ref{fig:error_dipole_T4}. Here, the variational polaron, weak phonon and polariton-polaron master equations all converge to roughly the same prediction accuracy as the light-coupling increases. This shows that the phonon-sideband effects that only the variational polaron master equation can capture are vanishing since the other master equations predict the emission spectra with an equal accuracy. The polaron master equation is seen to deviate marginally when the light-matter coupling increased, which is consistent with the previous observations.

\subsection{Discussion}
From the calculations performed in the Purcell regime and in the strong-coupling regime, it is evident that the variational approach is very versatile and performs well over a large range of light-matter coupling rates and also at elevated temperatures. The variationally-optimized master equation is therefore a good choice, with the notable exception of the intermediate regime where neither the light-matter coupling, nor the phonon coupling is dominating and therefore can be treated perturbatively. This is particularly visible in Fig. \ref{fig:error_T4}, where an increase in the relative error is seen when the light-matter coupling approaches the phonon cutoff frequency. Worth noting is also that the variational approach appears to be more numerically stable than the standard polaron and weak phonon master equations. In particular, numerical problems are encountered for the weak phonon master equation at elevated temperatures, where small positive real parts of the eigenvalues can lead to pronounced artifacts.

In the strong-coupling regime, the polariton-polaron approach is superior. By describing the system in the basis of the polaritons, the perturbation strength associated with the phonons is reduced which allows for a more precise inclusion of phonons. Also, most importantly, the transformation modifies the cavity operator so that non-Markovian effects such as the polariton-polaron sideband can be included through the cavity operator approach.


\section{Estimation of perturbation strength}
The different basis transformations employed in the master equations imply that the perturbative terms differ in terms of physics as well as magnitude. In this section we evaluate the magnitude of the effective perturbation strength, as this provides a measure of the accuracy to be expected from the corresponding master equation. The strength of the perturbation can be estimated from the terms making up the phononic dissipator as~\cite{dara_phd}:
\begin{equation}
    \int_0^\infty \mathrm{d}\tau \sum_{ij} |C_{ij}(\tau) A_i A_j(-\tau)|.
\end{equation}
The phonon correlation functions $C_{ij}(\tau)$ achieve their maximum value for $\tau = 0$ and decay on a timescale of $\nu_c$. The integral can therefore be estimated roughly as $g_{ij}\frac{|C_{ij}(0)|}{\nu_c}$, with $g_{ij}$ being the contribution of the operators $A_i$ and $A_j(-\tau)$ and are of the following form $g_{XX} = g_{YY} = g^2$ and $g_{ZY} = g_{YZ} = g$ and $g_{ZZ} = 1$ \cite{dara_phd}. This perturbative term is required to be small compared to higher order terms, to ensure the accuracy of the master equation. Since $\expval{H_I} = 0$, the next term in the master equation expansion is of fourth order and is estimated as $g_{ij}^2 \frac{|C_{ij}(0)|^2}{\nu_c^3}$ \cite{dara_phd}. This leads to the following condition for the validity of the master equation:
\begin{equation}
    PS = \frac{\sum_{ij} g_{ij}^2|C_{ij}(0)|^2}{\nu_c^2\sum_{ij} g_{ij}|C_{ij}(0)|} \ll 1.
\end{equation}
\begin{figure}
    \centering
    \includegraphics[width=0.99\linewidth]{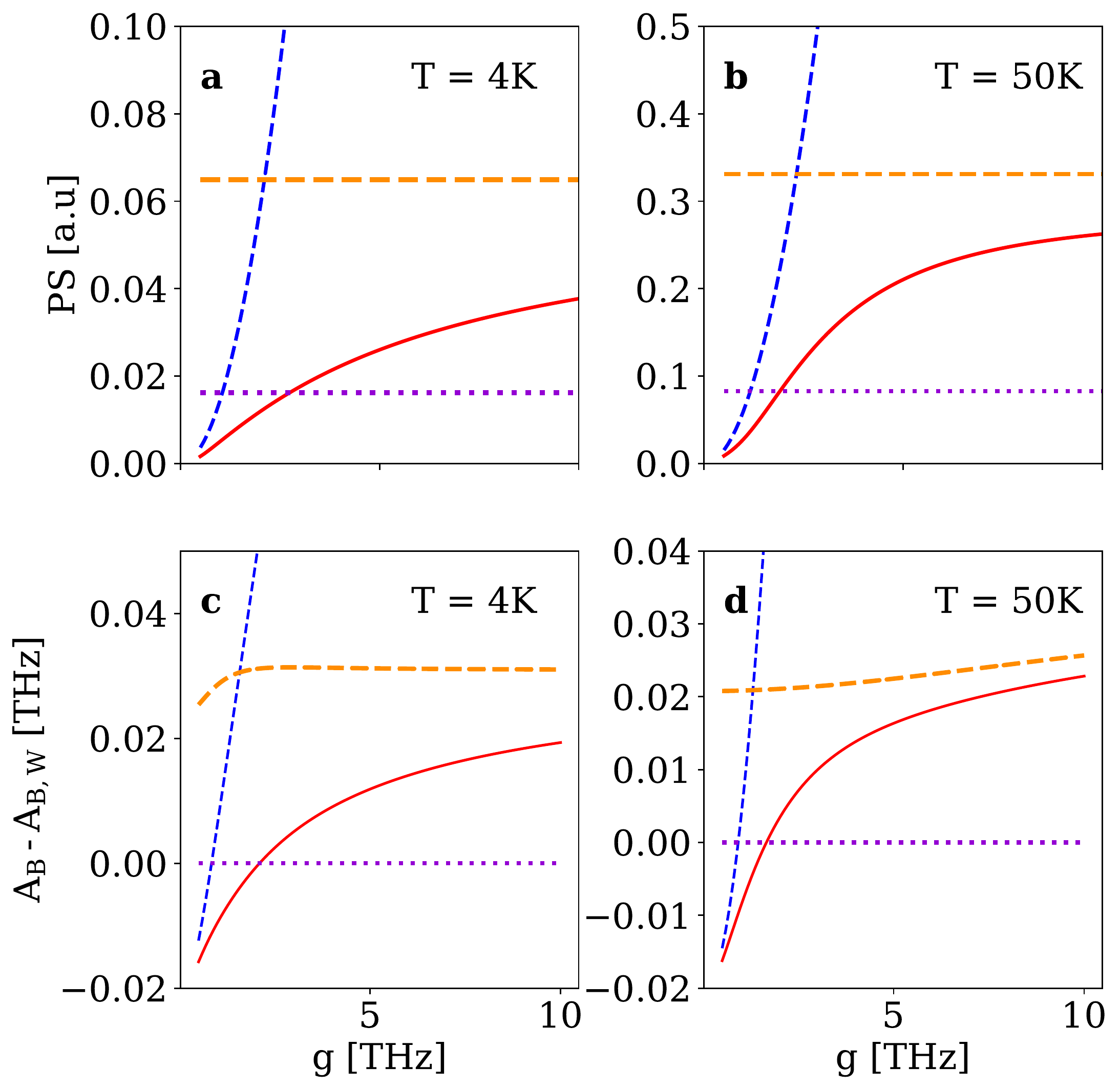}
    \caption{a and b: The strength of the second order perturbation terms in the master equations as function of the light-matter coupling for $\kappa = 0.5 \ \mathrm{THz}$ at (a) $T=4 \ \mathrm{K}$ and (b) $T=50\ \mathrm{K}$. c and d: The upper bound on the free energy at (c) $T=4 \ \mathrm{K}$ and (d) $T=50\ \mathrm{K}$. The upper bound on free energy of the polariton-polaron approach $A_{\rm B,W}$ has been subtracted from all bounds to enhance the differences. Linestyles are the same as in Figs. \ref{fig:cav_emission} and \ref{fig:p4_emission}.}
    \label{fig:strength_perturb_t4}
\end{figure}
The resulting perturbation strengths $PS$ as function of the light-matter coupling can be seen for $T=4 \ \mathrm{K}$ and $T=50 \ \mathrm{K}$ in Fig. \ref{fig:strength_perturb_t4}a and b, respectively. Noticeably, the perturbation parameter of the polariton-polaron approach is significantly lower than for the other approaches in the strong light-matter coupling regime. This emphasizes the fact that the polariton-polaron transformed basis states are closer to the true eigenstates of the coupled system in this regime. The standard polaron approach is also seen to imply a very high perturbation strength in the strong light-matter coupling regime which explains its failures in the emission spectra.
The variational polaron approach has a lower perturbation strength than the standard polaron and weak phonon approach for all light matter couplings. This is as expected since it was optimized to reduce the perturbative strength. The variational polaron transformation does, however, have limitations since the displacement of phonons depends only on the exciton state. The polariton-polaron approach, that displaces the phonon modes depending on the polariton state, captures the strong light-matter coupling dynamics better and therefore has a lower perturbation strength in this regime.

The fact that the polariton-polaron transformation leads to a basis closer to the true eigenstates of the coupled system is further supported by considering the upper bound on the free energy seen in Fig. \ref{fig:strength_perturb_t4}c and d. The polariton-polaron has a lower Bogoliubov bound on the free energy than the variational polaron transformation in the strong light-matter coupling regime. Thus, the variational polaron transformation minimized the upper bound on the free energy given the constraints of the transformation. However, applying the polariton-polaron transformation leads to a lower upper bound on the free energy, indicating that diagonality in the excitonic basis does not lead to a globally optimal transformation.


\section{Discussion of non-Markovian effects}
As stated in Sec.~\ref{sec:intro}, the non-Markovian response of the environment is particularly pronounced in the two-time correlation functions of the system, as compared to one-time expectation values. Thus, the emission spectrum, which relies on the evaluation of two-time correlation functions, is sensitive to the accuracy of the four perturbative master equation approaches, and their ability to capture non-Markovian effects. In this section, we show that the discrepancy between the perturbative approaches is far less pronounced, when one-time averages are used for benchmarking.

%
%
\begin{figure}[t]
    \centering
    \includegraphics[width=0.99\linewidth]{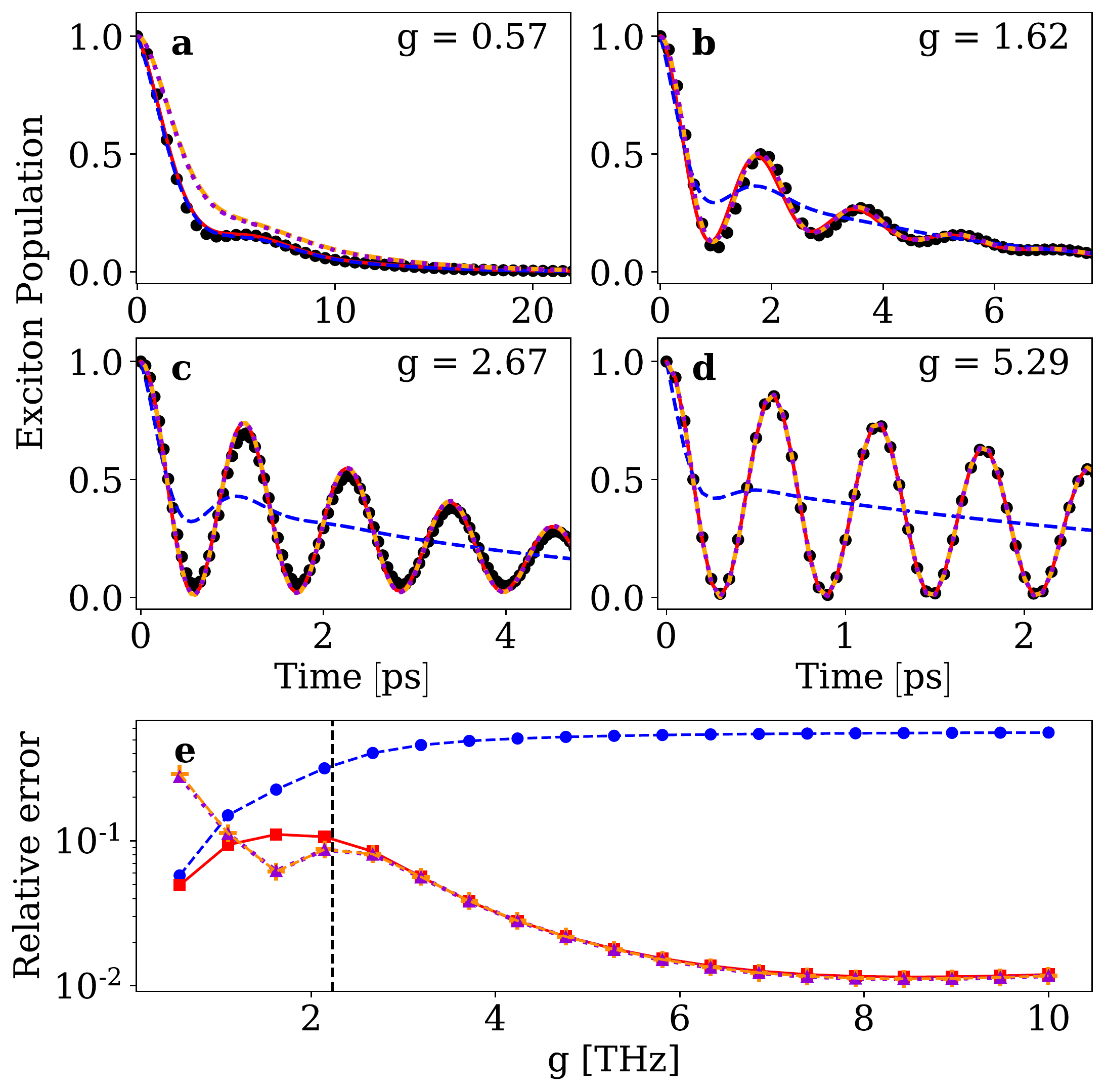}
    \caption{a-d: Time evolution of exciton population as predicted by the master equations and the tensor-network for varying light-matter coupling $g$, $\kappa = 0.5 \ \mathrm{THz}$ and at $T=50\ \mathrm{K}$. e: The corresponding relative error of the exciton population, computed in the same manner as eq. \eqref{eq:rel_error}. Linestyles are the same as in fig. \ref{fig:cav_emission} and \ref{fig:p4_emission}.}
    \label{fig:dynamics_compare_T150}
\end{figure}
In Fig. \ref{fig:dynamics_compare_T150}, the time evolution of the exciton population is shown for $T=50 \ \mathrm{K}$. In Fig. \ref{fig:dynamics_compare_T150}a, for $g=0.57 \ \mathrm{THz}$, the polariton-polaron and the weak phonon approaches both deviate clearly from the tensor-network result, which was also the case in the emission spectrum. As the light-matter coupling increases, the standard polaron approach breaks down as expected, leading to nonsensical results. The variational polaron, weak phonon and polariton-polaron approaches all converge for light-matter coupling rates exceeding the phonon cutoff frequency and predict the emitter evolution very accurately. This is also illustrated in Fig. \ref{fig:dynamics_compare_T150}e where the relative error converges to around $1 \%$. In comparison, Fig.~\ref{fig:error_T4} shows a relative error in the spectrum of $16\%$ for the weak phonon-coupling and variational polaron approaches, while the polariton-polaron approach only showed $2 \%$ deviation.  For strong light-matter couplings, the three master equation approaches are thus almost indistinguishable with regards to the population dynamics, but very different when considering the emission spectrum. As mentioned, this stems from the fact that the non-Markovian effects such as the polariton-polaron sidebands primarily enter through the multi-time correlation function rather than one-time expectation values.



\section{Conclusion}
In conclusion, we have introduced a new master equation, the polariton-polaron master equation, which is valid when a localized exciton state is coupled to a cavity mode with a light-matter coupling rate that exceeds the typical vibrational frequency of the environment. The master equation captures non-Markovian features such as the phonon dressing of the individual polariton peaks, which is not accounted for by previously formulated master equations. The polariton-polaron master equation was benchmarked with a numerically convergent tensor-network, showing that it indeed leads to very accurate results in the strong light-matter coupling regime.

A variationally optimized polaron master equation was also introduced \cite{nazir2016modelling}. Although less accurate than the polariton-polaron approach in the strong-coupling regime, it was shown to give accurate results over a large span of light-matter coupling rates with both high and low cavity decay rates, highlighting the versatility of this approach. The benchmarks made in this paper can serve as general guidelines to choosing the optimal master equation when considering an exciton-cavity system coupled to a vibrational environment. These results and observations are of a general nature, depending only on the relative magnitude of the phonon cutoff frequency and the light-matter coupling rate, and are thus applicable to other systems. Other such platforms include Transition metal dichalcogenides \cite{kleemann2017strong,geisler2019single,qin2020revealing}, Single methylene blue molecules \cite{chikkaraddy2016single} and nitrogen-vacancy centers \cite{NV_center1,NV_center2}.



\section*{Acknowledgment}
This work was supported by the Danish National Research Foundation through NanoPhoton - Center for Nanophotonics, Grant No. DNRF147 and Villum Fonden through the NATEC center (grant 8692).
EVD acknowledges support from Independent Research Fund Denmark through an International Postdoc fellowship.

\renewcommand\thefigure{\thesection.\arabic{figure}}
\appendix
\setcounter{figure}{0}
\section{Polariton-Polaron transformation \label{app:Polariton_polaron}}
In this appendix, we describe the details of the polariton-polaron transformation. We consider the Hamiltonian in the basis of the polaritons, Eq. \eqref{eq:polariton_Hamiltonian}, and apply the unitary transformation $WHW^\dagger$ defined in Eq. \eqref{eq:polariton_transformation}. The transformation has the following effect on the operators making up the system part of the Hamiltonian:
\begin{align}
 W^\dagger p^\dagger p W &= p^\dagger p \\
 W^\dagger m^\dagger m W &= m^\dagger m \\
 W^\dagger p^\dagger m W &= p^\dagger m \mathrm{e}^{Q} \\
 W^\dagger m^\dagger p W &= m^\dagger p \mathrm{e}^{-Q},
\end{align}
where
\begin{align}
Q=\sum_\textbf{k} \frac{f_\textbf{k}^+-f_\textbf{k}^-}{\nu_\textbf{k}}(b_\textbf{k}^\dagger - b_\textbf{k})
\end{align}
The transformation of the phonon operators is found using the displacement transformation~\cite{manyparticle}:
\begin{align*}
\begin{split}
& D_\textbf{k}\left(\frac{f_\textbf{k}}{\nu_\textbf{k}}\right) b_\mathbf{k} \:D_\textbf{k}\left(- \frac{f_\textbf{k}}{\nu_\textbf{k}}\right) = b_\mathbf{k}  - \frac{f_\mathbf{k}}{\nu_\mathbf{k}},
\end{split}
\end{align*}
which gives
\begin{align}
Wb_\textbf{k}W^\dagger = b_\mathbf{k} - \frac{f_\mathbf{k}^+}{\nu_\mathbf{k}}p^\dagger p -\frac{f_\mathbf{k}^-}{\nu_\mathbf{k}}m^\dagger m.
\end{align}
Inserting this and rearanging terms leads to the transformed Hamiltonian:
\begin{align}
\label{eq:polaron-polariton-transformed-1}
\begin{split}
    &W H W^\dagger = E_+ p^\dagger p + E_- m^\dagger m + \sum_\mathbf{k} \left[ \hbar\nu_\mathbf{k}b_\mathbf{k}^\dagger b_\mathbf{k} \vphantom{\frac{g_\textbf{k}}{f_\textbf{k+}}} \right. \\ &   - \hbar f_\mathbf{k}^+p^\dagger p \left( b_\mathbf{k}^\dagger + b_\mathbf{k} - C_+^2\frac{g_\mathbf{k}}{f_\mathbf{k}^+} b_\mathbf{k}^\dagger - C_+^2\frac{g_\mathbf{k}}{f_\mathbf{k}^+} b_\mathbf{k} \right) \\ & - \hbar f_\mathbf{k}^- m^\dagger m \left(  b_\mathbf{k}^\dagger + b_\mathbf{k} - C_-^2\frac{g_\mathbf{k}}{f_\mathbf{k}^-}b_\mathbf{k}^\dagger - C_-^2\frac{g_\mathbf{k}}{f_\mathbf{k}^-}b_\mathbf{k} \right) \\ & + \frac{\hbar {f_\mathbf{k}^+}^2}{\nu_\mathbf{k}} p^\dagger p  \left(1 - 2 \frac{C_+^2 g_\mathbf{k}}{f_\mathbf{k}^+} \right) + m^\dagger m \frac{\hbar {f_\mathbf{k}^-}^2}{\nu_\mathbf{k}} \left( 1 - 2 \frac{C_-^2 g_\mathbf{k}}{f_\mathbf{k}^-}\right)\\ & - C_+C_- \hbar g_\mathbf{k} \mathrm{e}^{Q} p^\dagger m\left(b_\mathbf{k}^\dagger + b_\mathbf{k}\right)  - C_+C_- \hbar g_\mathbf{k} \mathrm{e}^{-Q} m^\dagger p\left(b_\mathbf{k}^\dagger + b_\mathbf{k}\right) \\  & + \left. \frac{2 C_+C_- \hbar g_\mathbf{k}}{\nu_\mathbf{k}} \left( f_\mathbf{k}^-p^\dagger m \mathrm{e}^{Q} + f_\mathbf{k}^+m^\dagger p \mathrm{e}^{-Q} \right) \right]
\end{split}
\end{align}
In similar fashion to the standard polaron transformation, the coefficients $f_\textbf{k}^\pm$ are chosen so that the interaction terms that are diagonal in the dressed state basis, i.e. the second and third lines in Eq.~\eqref{eq:polaron-polariton-transformed-1}, vanish. This amounts to setting $f_\textbf{k}^\pm = C_\pm^2 g_\textbf{k}$. Thereby, the transformed Hamiltonian reduces to:
\begin{align}
\begin{split}
    &W H W^\dagger = E_+ p^\dagger p + E_- m^\dagger m \vphantom{\frac{g_\textbf{k}}{f_\textbf{k+}}} \\ & +  \sum_\mathbf{k} \left[ \hbar\nu_\mathbf{k}b_\mathbf{k}^\dagger b_\mathbf{k} - C_+^4\frac{\hbar g_\textbf{k}^2}{\nu_\mathbf{k}} p^\dagger p - C_-^4\frac{\hbar g_\textbf{k}^2}{\nu_\mathbf{k}} m^\dagger m \right. \\ & - C_+C_- \hbar g_\mathbf{k} (b_\mathbf{k}^\dagger + b_\mathbf{k})\left( \mathrm{e}^{Q} p^\dagger m + \mathrm{e}^{-Q} m^\dagger p\right) \\  & + \left. \frac{2 C_+C_- \hbar g_\mathbf{k}^2}{\nu_\mathbf{k}} \left( C_-^2 p^\dagger m \mathrm{e}^{Q} + C_+^2 m^\dagger p \mathrm{e}^{-Q} \right) \right]
\end{split}
\end{align}
As stated earlier, the detuning is assumed to be zero, $\Delta = 0$, which implies that $C_+ = C_- = 1/\sqrt{2}$ and therefore also $f_\textbf{k+} = f_\textbf{k-}$ and $\mathrm{e}^Q = \mathrm{e}^{-Q} = 1$. This reduces the complexity significantly and introducing the phonon-shift $\Delta_p = -\sum_\textbf{k}\frac{g_\textbf{k}^2}{\nu_\textbf{k}}$ leads to the Hamiltonian in eq. \eqref{eq:polariton_Hamiltonian}.

\section{Tensor Network \label{app:Tensor_network}}
\begin{figure}
    \centering
    \includegraphics[width=\linewidth]{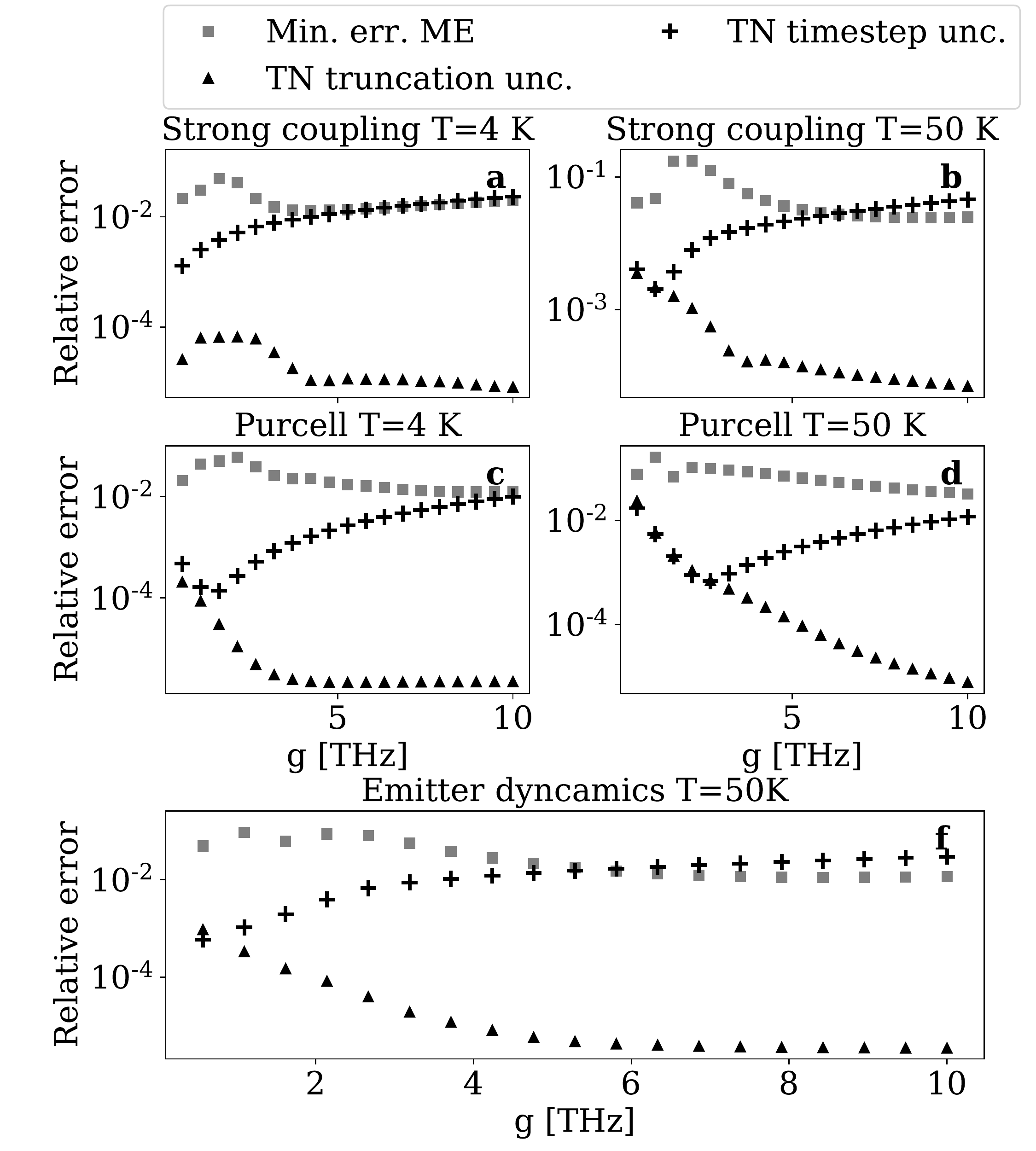}
    \caption{The estimated accuracy of the tensor-network shown as the uncertainty due to finite truncation error (black triangles) and the uncertainty due to finite number of time steps (black crosses), presented together with the lowest relative error of the four master equations at each value of the light-matter coupling strength, $g$. The different panels relate to the different error plots shown throughout the paper and have the same parameters as the figure they relate to. a-b: Relates to Fig. \ref{fig:error_T4}. c-d: Relates to Fig. \ref{fig:error_dipole_T4}. e: Relates to Fig. \ref{fig:dynamics_compare_T150}e.}
    \label{fig:tensor_accuracy}
\end{figure}

\begin{table}
    \centering
    \begin{tabular}{l|c|c|c}
    Configuration, temperature & $\lambda_{\rm c}$ & $t_{\rm max}$ & $N_t$ \\ \hline
    Strong coupling, 4K & $5\times 10^{-9}$ & 150 ps & 3000 \\
    Strong coupling, 50K & $5\times 10^{-9}$ & 150 ps & 3000 \\
    Purcell, 4K & $2.5\times 10^{-9}$ & 60 ps & 3000 \\
    Purcell, 50K & $2.5\times 10^{-9}$ & 50 ps & 3000 \\ \hline
    \end{tabular}
    \caption{Numerical parameters used for the tensor-network reference calculations. The configuration 'strong coupling' refers to the parameter setting with $\kappa$ fixed at $0.5\mathrm{\:ps^{-1}}$ (used in Figs.~\ref{fig:cav_emission}, \ref{fig:error_T4} and \ref{fig:unphysical}), and 'Purcell' refers to the setting with $\kappa$ pinned to $4g$ (used in Figs.~\ref{fig:p4_emission} and \ref{fig:error_dipole_T4}).}
    \label{tab:tensor-parameters}
\end{table}

The tensor network algorithm used for reference calculations of the spectrum and dynamics was developed in Refs.~\onlinecite{Emil_tensor1,Emil_tensor2}, and the implementation for the particular system studied in this paper is presented in Ref.~\cite{Denning_2020_phonon_decoupling}. The numerical accuracy in the calculated spectrum, $S_{\rm tn}(\omega)$, or exciton population dynamics, $P_{\rm tn}(t)$, can be traced back to two numerical parameters, namely the singular-value truncation cutoff~\cite{Emil_tensor1,Emil_tensor2}, $\lambda_{\rm c}$ (given relative to the maximum singular value in each compression step), and the timestep, $\delta_t=t_{\rm max}/N_t$, where $t_\mathrm{max}$ is the total time span of the calculation and $N_t$ is the number of time points used for time discretization. The numerical parameters used for the reference calculations are presented in Table.~\ref{tab:tensor-parameters}.

For each calculation of the spectrum or the population dynamics, the numerical convergence with respect to the truncation error and time discretization is studied in order to ensure that the algorithm is operating in a numerically convergent regime. Furthermore, we also used the convergence to estimate the accuracy of the tensor-network reference calculations. Using the symbol $X(x;\lambda_{\rm c}, N_t)$ to denote either the spectrum $S_{\rm tn}(\omega)$ (where $x=\omega$) or exciton population dynamics $P_{\rm tn}(t)$ (where $x=t$) calculated with truncation error $\lambda_{\rm c}$ and $N_t$ time discretization points, the convergence properties are evaluated through the relative deviation function defined as
\begin{align}
\begin{split}
    \Delta X(&\lambda_{\rm c}^{(1)},N_t^{(1)};\lambda_{\rm c}^{(2)}, N_t^{(2)}) \\ &= \qty(\frac{\int\dd{x} [X(x;\lambda_{\rm c}^{(1)}, N_t^{(1)})- X(x;\lambda_{\rm c}^{(2)}, N_t^{(2)})]^2  }{\int\dd{x} X(x,\lambda_{\rm c}^{(1)} N_t^{(1)})^2})^{1/2}.
\end{split}
\end{align}
The uncertainty due to the finite truncation error $\lambda_{\rm c}$ is then estimated by comparing to a calculation with $2\lambda_{\rm c}$, i.e. by evaluating the deviation $\Delta X(\lambda_{\rm c},N_t;2\lambda_{\rm c},N_t)$. This uncertainty is shown in Fig.~\ref{fig:tensor_accuracy} for all calculations presented in the article (black triangle markers). For comparison, the smallest benchmark error obtained from the master equations is also shown (grey square markers)). Similarly, the uncertainty due to finite time step is estimated by evaluating the deviation $\Delta X(\lambda_{\rm c},N_t;\lambda_{\rm c},N_t/2)$, shown with black cross markers in Fig.~\ref{fig:tensor_accuracy}. In the parameter regions where the truncation error uncertainty is far below the time-step uncertainty, it is clear that the accuracy of the calculation can be well approximated by the time-step uncertainty, because the truncation error does not contribute with any appreciable uncertainty. In the regions where the errors are comparable, we assess that this is still the case: Since uncertainty due to finite truncation error lead to variations between the two influence functionals calculated with $N_t$ and $N_t/2$ time steps, the truncation error is also represented in the time step uncertainty $\Delta X(\lambda_{\rm c},N_t;\lambda_{\rm c},N_t/2)$. This effect is also observable in Fig.~\ref{fig:tensor_accuracy}b-d (most noticably in Fig.~\ref{fig:tensor_accuracy}d), where the time step uncertainty is consistently higher than the truncation error uncertainty.

We note that there are parameter regions, where the estimated uncertainty of the tensor-network reference calculations exceeds or is very close to the deviation of the best master equation. In these regions, the accuracy of the reference calculations and the deviation of the best-performing master equation are both on the order of 1\%, meaning that the master equation is performing with accuracy within 1\% of the true result.

Due to technical challenges with the convergence of the singular-value decomposition at $T=4\mathrm{\:K}$ with $N_t=3000$ and $t_{\rm max}=50\mathrm{\:ps}$, we have made an exception in the calculation of the time step uncertainty for these calculations. Here, the deviation $\Delta X(\lambda_{\rm c},3000;\lambda_{\rm c}, 1500)$ is calculated with $t_{\rm max}=60\mathrm{\:ps}$ for $N_t=3000$ and with $t_{\rm max}=50\mathrm{\: ps}$ for $N_t=1500$.

\setcounter{figure}{0}
\section{Estimating the phonon-shift and renormalization \label{app:split_and_shift}}
\begin{figure}[t]
    \centering
    \includegraphics[width = \linewidth]{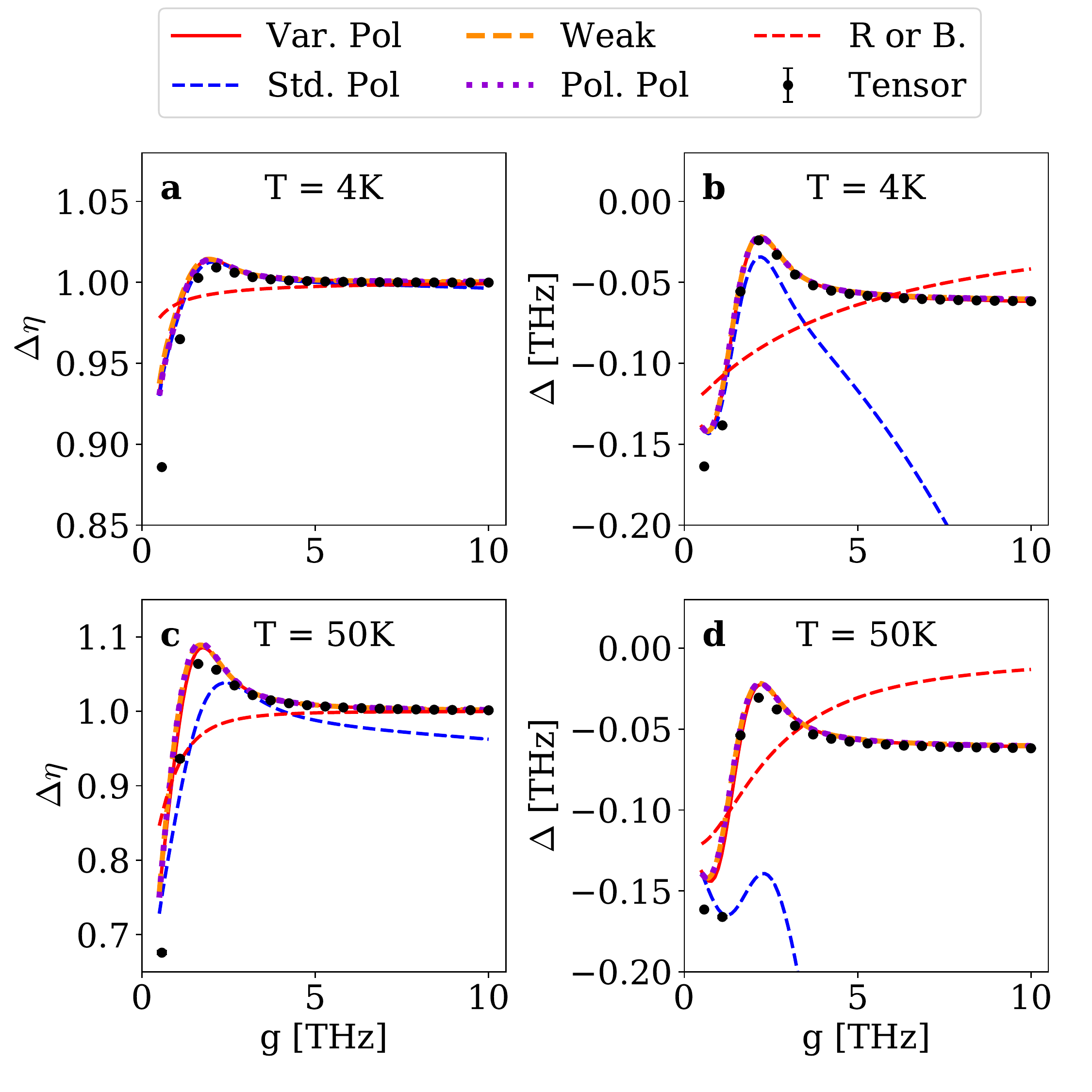}
    \caption{a and c: The renormalization of the light-matter coupling rate calculated from equation \eqref{eq:renorm} as a function of $g$ at (a) $T=4 \ \mathrm{K}$ and (c) $T=50 \ \mathrm{K}$. b and d: The estimated phonon shift calculated from the positions of polariton peaks (b) $T=4 \ \mathrm{K}$ and (d) $T=50 \ \mathrm{K}$. The peak positions have been extracted from the eigenvalues of the Liouvillian for the four master equations and by fitting a Lorentzian curve to the tensor-network results. The error bars indicate the uncertainty of the peak splitting and shift in the tensor-network data, evaluated by comparing spectra calculated with different truncation cutoff and time discretization in analogy with the error estimation presented in Appendix~\ref{app:Tensor_network}.}
    \label{fig:split_shift_t4}
\end{figure}

The variational polaron transformation introduces a phonon induced shift $R$ to the emitter frequency $\omega_{eg}$, as given by equation \eqref{eq:R_V}. The naive guess for the frequency shift due to phonons would therefore be $R$, since it explicitly changes the emitter frequency in the Hamiltonian. However, a second-order phonon-induced Lamb-shift originating from the phononic-dissipator $\mathcal{K}[\rho]$ constitutes a second non-negligible effect on the phonon shift. To extract the actual emitter shift due to phonons, the frequencies of the polariton peaks are computed by considering the eigenvalues of the Liouvillian. The total shift can be estimated as $S_+ + S_-$, where $S_\pm$ denotes the frequency of the upper/lower polariton peak. According to the Jaynes-Cummings model, the eigenvalues and thus the positions of the polariton peaks are: $E_\pm = \frac{\Delta}{2} \pm \sqrt{4g^2 + \Delta^2}/2$, where $\Delta$ is the detuning of the system. In our case, the detuning without the inclusion of phonons is 0 and thus $E_+ + E_- = \Delta$ is the detuning solely imparted by the phonons. To extract the polariton peak frequency predicted by the tensor-network, we fit a Lorentzian curve to each peak. In fig. \ref{fig:split_shift_t4}b and d, the naive guess $R$ is shown together with the actual shift predicted by the tensor-network and the shift found by studying the eigenvalues. It is clear that the value of the shift $R$ does not constitute a good approximation of the emitter-shift due to phonons. By including the second-order phonon-induced Lamb-shift, found from the eigenvalues, the variations with $g$ of the renormalization and the detuning can be well explained by the variational polaron, weak coupling and polariton-polaron master equations. On the other hand, the standard polaron master equation fails to capture these effects at elevated temperatures and high light-matter coupling rates as was expected. Estimating the resonance frequency of the system can thus be done by using either the variational polaron, weak coupling or polariton-polaron master equation.

Phonons also give rise to a change to the effective light-matter coupling strength. This is illustrated explicitly by the renormalization factor $B$ from the variational polaron transformation given in eq. \eqref{eq:R_B}. In a similar fashion to the phonon shift, there is also a contribution to the renormalization from second-order effects introduced by the phonon-dissipator. These second-order effects can not be neglected either, as illustrated in fig. \ref{fig:split_shift_t4}a and c, where the light-matter coupling renormalization is shown. The renormalization has here been estimated as:
\begin{equation}
    \Delta \eta = \frac{\sqrt{(S_+ - S_-)^2-(S_+ + S_-)^2}}{2g} \label{eq:renorm}
\end{equation}
where the numerator, by insertion of the Jaynes-Cummings eigenvalues, gives $2g'$, where $g'$ is the effective light-matter coupling. Thus, by normalizing with $2g$ one finds the renormalization-factor of the light-matter coupling. Again, the naive first order guess, $B$, as the renormalization factor is seen to be deviate from the actual renormalization. Estimating the renormalization factor by including second order effects, all the master equations, except the standard polaron master equation, predict the renormalization well, however. The renormalization factor converges to one, and is thus only important for small light-matter coupling rates, $g$. It is worth noting that the prediction of $\Delta \eta > 1$ is a robust prediction and determining it via other means such as a fit on the evolution of the emitter-population yields similar values. Other numerical studies also show similar results which further validates the prediction \cite{Phonon_dephasing_Muljarov}. The implication of $\Delta \eta > 1$ is that phonons enhance the light-matter coupling which is opposite to what is normally expected from phonons, i.e., the renormalization factor $B$ is always smaller than 1.


\begin{thebibliography}{74}%
\makeatletter
\providecommand \@ifxundefined [1]{%
 \@ifx{#1\undefined}
}%
\providecommand \@ifnum [1]{%
 \ifnum #1\expandafter \@firstoftwo
 \else \expandafter \@secondoftwo
 \fi
}%
\providecommand \@ifx [1]{%
 \ifx #1\expandafter \@firstoftwo
 \else \expandafter \@secondoftwo
 \fi
}%
\providecommand \natexlab [1]{#1}%
\providecommand \enquote  [1]{``#1''}%
\providecommand \bibnamefont  [1]{#1}%
\providecommand \bibfnamefont [1]{#1}%
\providecommand \citenamefont [1]{#1}%
\providecommand \href@noop [0]{\@secondoftwo}%
\providecommand \href [0]{\begingroup \@sanitize@url \@href}%
\providecommand \@href[1]{\@@startlink{#1}\@@href}%
\providecommand \@@href[1]{\endgroup#1\@@endlink}%
\providecommand \@sanitize@url [0]{\catcode `\\12\catcode `\$12\catcode
  `\&12\catcode `\#12\catcode `\^12\catcode `\_12\catcode `\%12\relax}%
\providecommand \@@startlink[1]{}%
\providecommand \@@endlink[0]{}%
\providecommand \url  [0]{\begingroup\@sanitize@url \@url }%
\providecommand \@url [1]{\endgroup\@href {#1}{\urlprefix }}%
\providecommand \urlprefix  [0]{URL }%
\providecommand \Eprint [0]{\href }%
\providecommand \doibase [0]{http://dx.doi.org/}%
\providecommand \selectlanguage [0]{\@gobble}%
\providecommand \bibinfo  [0]{\@secondoftwo}%
\providecommand \bibfield  [0]{\@secondoftwo}%
\providecommand \translation [1]{[#1]}%
\providecommand \BibitemOpen [0]{}%
\providecommand \bibitemStop [0]{}%
\providecommand \bibitemNoStop [0]{.\EOS\space}%
\providecommand \EOS [0]{\spacefactor3000\relax}%
\providecommand \BibitemShut  [1]{\csname bibitem#1\endcsname}%
\let\auto@bib@innerbib\@empty
\bibitem [{\citenamefont {O'Brien}\ \emph {et~al.}(2009)\citenamefont
  {O'Brien}, \citenamefont {Furusawa},\ and\ \citenamefont
  {Vučković}}]{O_brian_semiconductor_2009}%
  \BibitemOpen
  \bibfield  {author} {\bibinfo {author} {\bibfnamefont {J.~L.}\ \bibnamefont
  {O'Brien}}, \bibinfo {author} {\bibfnamefont {A.}~\bibnamefont {Furusawa}}, \
  and\ \bibinfo {author} {\bibfnamefont {J.}~\bibnamefont {Vučković}},\
  }\href@noop {} {\bibfield  {journal} {\bibinfo  {journal} {Nature Photonics}\
  }\textbf {\bibinfo {volume} {3}},\ \bibinfo {pages} {687} (\bibinfo {year}
  {2009})}\BibitemShut {NoStop}%
\bibitem [{\citenamefont {Iles-Smith}\ \emph
  {et~al.}(2017{\natexlab{a}})\citenamefont {Iles-Smith}, \citenamefont
  {McCutcheon}, \citenamefont {Nazir},\ and\ \citenamefont
  {M{\o}rk}}]{simultanoeusefficiency}%
  \BibitemOpen
  \bibfield  {author} {\bibinfo {author} {\bibfnamefont {J.}~\bibnamefont
  {Iles-Smith}}, \bibinfo {author} {\bibfnamefont {D.}~\bibnamefont
  {McCutcheon}}, \bibinfo {author} {\bibfnamefont {A.}~\bibnamefont {Nazir}}, \
  and\ \bibinfo {author} {\bibfnamefont {J.}~\bibnamefont {M{\o}rk}},\ }\href
  {\doibase 10.1038/NPHOTON.2017.101} {\bibfield  {journal} {\bibinfo
  {journal} {Nature Photonics}\ }\textbf {\bibinfo {volume} {11}},\ \bibinfo
  {pages} {521} (\bibinfo {year} {2017}{\natexlab{a}})}\BibitemShut {NoStop}%
\bibitem [{\citenamefont {Carmele}\ and\ \citenamefont
  {Reitzenstein}(2019)}]{non_markovian_QD}%
  \BibitemOpen
  \bibfield  {author} {\bibinfo {author} {\bibfnamefont {A.}~\bibnamefont
  {Carmele}}\ and\ \bibinfo {author} {\bibfnamefont {S.}~\bibnamefont
  {Reitzenstein}},\ }\href {\doibase https://doi.org/10.1515/nanoph-2018-0222}
  {\bibfield  {journal} {\bibinfo  {journal} {Nanophotonics}\ }\textbf
  {\bibinfo {volume} {8}},\ \bibinfo {pages} {655 } (\bibinfo {year}
  {2019})}\BibitemShut {NoStop}%
\bibitem [{\citenamefont {Besombes}\ \emph
  {et~al.}(2001{\natexlab{a}})\citenamefont {Besombes}, \citenamefont {Kheng},
  \citenamefont {Marsal},\ and\ \citenamefont
  {Mariette}}]{besombes2001acoustic}%
  \BibitemOpen
  \bibfield  {author} {\bibinfo {author} {\bibfnamefont {L.}~\bibnamefont
  {Besombes}}, \bibinfo {author} {\bibfnamefont {K.}~\bibnamefont {Kheng}},
  \bibinfo {author} {\bibfnamefont {L.}~\bibnamefont {Marsal}}, \ and\ \bibinfo
  {author} {\bibfnamefont {H.}~\bibnamefont {Mariette}},\ }\href@noop {}
  {\bibfield  {journal} {\bibinfo  {journal} {Physical Review B}\ }\textbf
  {\bibinfo {volume} {63}},\ \bibinfo {pages} {155307} (\bibinfo {year}
  {2001}{\natexlab{a}})}\BibitemShut {NoStop}%
\bibitem [{\citenamefont {Brash}\ \emph {et~al.}(2019)\citenamefont {Brash},
  \citenamefont {Iles-Smith}, \citenamefont {Phillips}, \citenamefont
  {McCutcheon}, \citenamefont {O'Hara}, \citenamefont {Clarke}, \citenamefont
  {Royall}, \citenamefont {Wilson}, \citenamefont {M{\o}rk}, \citenamefont
  {Skolnick}, \citenamefont {Fox},\ and\ \citenamefont
  {Nazir}}]{lightscattering}%
  \BibitemOpen
  \bibfield  {author} {\bibinfo {author} {\bibfnamefont {A.~J.}\ \bibnamefont
  {Brash}}, \bibinfo {author} {\bibfnamefont {J.}~\bibnamefont {Iles-Smith}},
  \bibinfo {author} {\bibfnamefont {C.~L.}\ \bibnamefont {Phillips}}, \bibinfo
  {author} {\bibfnamefont {D.~P.~S.}\ \bibnamefont {McCutcheon}}, \bibinfo
  {author} {\bibfnamefont {J.}~\bibnamefont {O'Hara}}, \bibinfo {author}
  {\bibfnamefont {E.}~\bibnamefont {Clarke}}, \bibinfo {author} {\bibfnamefont
  {B.}~\bibnamefont {Royall}}, \bibinfo {author} {\bibfnamefont {L.~R.}\
  \bibnamefont {Wilson}}, \bibinfo {author} {\bibfnamefont {J.}~\bibnamefont
  {M{\o}rk}}, \bibinfo {author} {\bibfnamefont {M.~S.}\ \bibnamefont
  {Skolnick}}, \bibinfo {author} {\bibfnamefont {A.~M.}\ \bibnamefont {Fox}}, \
  and\ \bibinfo {author} {\bibfnamefont {A.}~\bibnamefont {Nazir}},\
  }\href@noop {} {\bibfield  {journal} {\bibinfo  {journal} {Physical Review
  Letters}\ }\textbf {\bibinfo {volume} {123}},\ \bibinfo {pages} {167403}
  (\bibinfo {year} {2019})}\BibitemShut {NoStop}%
\bibitem [{\citenamefont {Koong}\ \emph {et~al.}(2019)\citenamefont {Koong},
  \citenamefont {Scerri}, \citenamefont {Rambach}, \citenamefont {Santana},
  \citenamefont {Park}, \citenamefont {Song}, \citenamefont {Gauger},\ and\
  \citenamefont {Gerardot}}]{koong2019fundamental}%
  \BibitemOpen
  \bibfield  {author} {\bibinfo {author} {\bibfnamefont {Z.-X.}\ \bibnamefont
  {Koong}}, \bibinfo {author} {\bibfnamefont {D.}~\bibnamefont {Scerri}},
  \bibinfo {author} {\bibfnamefont {M.}~\bibnamefont {Rambach}}, \bibinfo
  {author} {\bibfnamefont {T.~S.}\ \bibnamefont {Santana}}, \bibinfo {author}
  {\bibfnamefont {S.-I.}\ \bibnamefont {Park}}, \bibinfo {author}
  {\bibfnamefont {J.~D.}\ \bibnamefont {Song}}, \bibinfo {author}
  {\bibfnamefont {E.~M.}\ \bibnamefont {Gauger}}, \ and\ \bibinfo {author}
  {\bibfnamefont {B.~D.}\ \bibnamefont {Gerardot}},\ }\href@noop {} {\bibfield
  {journal} {\bibinfo  {journal} {Physical Review Letters}\ }\textbf {\bibinfo
  {volume} {123}},\ \bibinfo {pages} {167402} (\bibinfo {year}
  {2019})}\BibitemShut {NoStop}%
\bibitem [{\citenamefont {F{\"o}rstner}\ \emph {et~al.}(2003)\citenamefont
  {F{\"o}rstner}, \citenamefont {Weber}, \citenamefont {Danckwerts},\ and\
  \citenamefont {Knorr}}]{forstner2003phonon}%
  \BibitemOpen
  \bibfield  {author} {\bibinfo {author} {\bibfnamefont {J.}~\bibnamefont
  {F{\"o}rstner}}, \bibinfo {author} {\bibfnamefont {C.}~\bibnamefont {Weber}},
  \bibinfo {author} {\bibfnamefont {J.}~\bibnamefont {Danckwerts}}, \ and\
  \bibinfo {author} {\bibfnamefont {A.}~\bibnamefont {Knorr}},\ }\href@noop {}
  {\bibfield  {journal} {\bibinfo  {journal} {Physical Review Letters}\
  }\textbf {\bibinfo {volume} {91}},\ \bibinfo {pages} {127401} (\bibinfo
  {year} {2003})}\BibitemShut {NoStop}%
\bibitem [{\citenamefont {Ramsay}\ \emph
  {et~al.}(2010{\natexlab{a}})\citenamefont {Ramsay}, \citenamefont {Gopal},
  \citenamefont {Gauger}, \citenamefont {Nazir}, \citenamefont {Lovett},
  \citenamefont {Fox},\ and\ \citenamefont {Skolnick}}]{ramsay2010damping}%
  \BibitemOpen
  \bibfield  {author} {\bibinfo {author} {\bibfnamefont {A.~J.}\ \bibnamefont
  {Ramsay}}, \bibinfo {author} {\bibfnamefont {A.~V.}\ \bibnamefont {Gopal}},
  \bibinfo {author} {\bibfnamefont {E.~M.}\ \bibnamefont {Gauger}}, \bibinfo
  {author} {\bibfnamefont {A.}~\bibnamefont {Nazir}}, \bibinfo {author}
  {\bibfnamefont {B.~W.}\ \bibnamefont {Lovett}}, \bibinfo {author}
  {\bibfnamefont {A.~M.}\ \bibnamefont {Fox}}, \ and\ \bibinfo {author}
  {\bibfnamefont {M.~S.}\ \bibnamefont {Skolnick}},\ }\href@noop {} {\bibfield
  {journal} {\bibinfo  {journal} {Physical Review Letters}\ }\textbf {\bibinfo
  {volume} {104}},\ \bibinfo {pages} {017402} (\bibinfo {year}
  {2010}{\natexlab{a}})}\BibitemShut {NoStop}%
\bibitem [{\citenamefont {Ramsay}\ \emph
  {et~al.}(2010{\natexlab{b}})\citenamefont {Ramsay}, \citenamefont {Godden},
  \citenamefont {Boyle}, \citenamefont {Gauger}, \citenamefont {Nazir},
  \citenamefont {Lovett}, \citenamefont {Fox},\ and\ \citenamefont
  {Skolnick}}]{ramsay2010phonon}%
  \BibitemOpen
  \bibfield  {author} {\bibinfo {author} {\bibfnamefont {A.~J.}\ \bibnamefont
  {Ramsay}}, \bibinfo {author} {\bibfnamefont {T.~M.}\ \bibnamefont {Godden}},
  \bibinfo {author} {\bibfnamefont {S.~J.}\ \bibnamefont {Boyle}}, \bibinfo
  {author} {\bibfnamefont {E.~M.}\ \bibnamefont {Gauger}}, \bibinfo {author}
  {\bibfnamefont {A.}~\bibnamefont {Nazir}}, \bibinfo {author} {\bibfnamefont
  {B.~W.}\ \bibnamefont {Lovett}}, \bibinfo {author} {\bibfnamefont {A.~M.}\
  \bibnamefont {Fox}}, \ and\ \bibinfo {author} {\bibfnamefont {M.~S.}\
  \bibnamefont {Skolnick}},\ }\href@noop {} {\bibfield  {journal} {\bibinfo
  {journal} {Physical Review Letters}\ }\textbf {\bibinfo {volume} {105}},\
  \bibinfo {pages} {177402} (\bibinfo {year} {2010}{\natexlab{b}})}\BibitemShut
  {NoStop}%
\bibitem [{\citenamefont {Ramsay}\ \emph {et~al.}(2011)\citenamefont {Ramsay},
  \citenamefont {Godden}, \citenamefont {Boyle}, \citenamefont {Gauger},
  \citenamefont {Nazir}, \citenamefont {Lovett}, \citenamefont {Gopal},
  \citenamefont {Fox},\ and\ \citenamefont {Skolnick}}]{ramsay2011effect}%
  \BibitemOpen
  \bibfield  {author} {\bibinfo {author} {\bibfnamefont {A.~J.}\ \bibnamefont
  {Ramsay}}, \bibinfo {author} {\bibfnamefont {T.~M.}\ \bibnamefont {Godden}},
  \bibinfo {author} {\bibfnamefont {S.~J.}\ \bibnamefont {Boyle}}, \bibinfo
  {author} {\bibfnamefont {E.~M.}\ \bibnamefont {Gauger}}, \bibinfo {author}
  {\bibfnamefont {A.}~\bibnamefont {Nazir}}, \bibinfo {author} {\bibfnamefont
  {B.~W.}\ \bibnamefont {Lovett}}, \bibinfo {author} {\bibfnamefont {A.~V.}\
  \bibnamefont {Gopal}}, \bibinfo {author} {\bibfnamefont {A.~M.}\ \bibnamefont
  {Fox}}, \ and\ \bibinfo {author} {\bibfnamefont {M.~S.}\ \bibnamefont
  {Skolnick}},\ }\href@noop {} {\bibfield  {journal} {\bibinfo  {journal}
  {Journal of Applied Physics}\ }\textbf {\bibinfo {volume} {109}},\ \bibinfo
  {pages} {102415} (\bibinfo {year} {2011})}\BibitemShut {NoStop}%
\bibitem [{\citenamefont {Galego}\ \emph {et~al.}(2016)\citenamefont {Galego},
  \citenamefont {Garcia-Vidal},\ and\ \citenamefont
  {Feist}}]{Feist_2016_Supressing_photochemical}%
  \BibitemOpen
  \bibfield  {author} {\bibinfo {author} {\bibfnamefont {J.}~\bibnamefont
  {Galego}}, \bibinfo {author} {\bibfnamefont {F.}~\bibnamefont
  {Garcia-Vidal}}, \ and\ \bibinfo {author} {\bibfnamefont {J.}~\bibnamefont
  {Feist}},\ }\href@noop {} {\bibfield  {journal} {\bibinfo  {journal} {Nature
  Communications}\ }\textbf {\bibinfo {volume} {7}} (\bibinfo {year}
  {2016})}\BibitemShut {NoStop}%
\bibitem [{\citenamefont {Herrera}\ and\ \citenamefont
  {Spano}(2016)}]{Frank_2016_cavity_controllerd}%
  \BibitemOpen
  \bibfield  {author} {\bibinfo {author} {\bibfnamefont {F.}~\bibnamefont
  {Herrera}}\ and\ \bibinfo {author} {\bibfnamefont {F.~C.}\ \bibnamefont
  {Spano}},\ }\href {\doibase 10.1103/PhysRevLett.116.238301} {\bibfield
  {journal} {\bibinfo  {journal} {Physical Review Letters}\ }\textbf {\bibinfo
  {volume} {116}},\ \bibinfo {pages} {238301} (\bibinfo {year}
  {2016})}\BibitemShut {NoStop}%
\bibitem [{\citenamefont {Galego}\ \emph {et~al.}(2015)\citenamefont {Galego},
  \citenamefont {Garcia-Vidal},\ and\ \citenamefont
  {Feist}}]{galego2015cavity}%
  \BibitemOpen
  \bibfield  {author} {\bibinfo {author} {\bibfnamefont {J.}~\bibnamefont
  {Galego}}, \bibinfo {author} {\bibfnamefont {F.~J.}\ \bibnamefont
  {Garcia-Vidal}}, \ and\ \bibinfo {author} {\bibfnamefont {J.}~\bibnamefont
  {Feist}},\ }\href@noop {} {\bibfield  {journal} {\bibinfo  {journal}
  {Physical Review X}\ }\textbf {\bibinfo {volume} {5}},\ \bibinfo {pages}
  {041022} (\bibinfo {year} {2015})}\BibitemShut {NoStop}%
\bibitem [{\citenamefont {Denning}\ \emph
  {et~al.}(2020{\natexlab{a}})\citenamefont {Denning}, \citenamefont
  {Bundgaard-Nielsen},\ and\ \citenamefont
  {M\o{}rk}}]{Denning_2020_phonon_decoupling}%
  \BibitemOpen
  \bibfield  {author} {\bibinfo {author} {\bibfnamefont {E.~V.}\ \bibnamefont
  {Denning}}, \bibinfo {author} {\bibfnamefont {M.}~\bibnamefont
  {Bundgaard-Nielsen}}, \ and\ \bibinfo {author} {\bibfnamefont
  {J.}~\bibnamefont {M\o{}rk}},\ }\href {\doibase 10.1103/PhysRevB.102.235303}
  {\bibfield  {journal} {\bibinfo  {journal} {Phys. Rev. B}\ }\textbf {\bibinfo
  {volume} {102}},\ \bibinfo {pages} {235303} (\bibinfo {year}
  {2020}{\natexlab{a}})}\BibitemShut {NoStop}%
\bibitem [{\citenamefont {Hughes}\ \emph {et~al.}(2021)\citenamefont {Hughes},
  \citenamefont {Settineri}, \citenamefont {Savasta},\ and\ \citenamefont
  {Nori}}]{hughes2021resonant}%
  \BibitemOpen
  \bibfield  {author} {\bibinfo {author} {\bibfnamefont {S.}~\bibnamefont
  {Hughes}}, \bibinfo {author} {\bibfnamefont {A.}~\bibnamefont {Settineri}},
  \bibinfo {author} {\bibfnamefont {S.}~\bibnamefont {Savasta}}, \ and\
  \bibinfo {author} {\bibfnamefont {F.}~\bibnamefont {Nori}},\ }\href@noop {}
  {\bibfield  {journal} {\bibinfo  {journal} {ArXiv preprint ArXiv:2103.08670}\
  } (\bibinfo {year} {2021})}\BibitemShut {NoStop}%
\bibitem [{\citenamefont {Doan}\ \emph {et~al.}(2005)\citenamefont {Doan},
  \citenamefont {Cao}, \citenamefont {Thoai},\ and\ \citenamefont
  {Haug}}]{doan2005condensation}%
  \BibitemOpen
  \bibfield  {author} {\bibinfo {author} {\bibfnamefont {T.~D.}\ \bibnamefont
  {Doan}}, \bibinfo {author} {\bibfnamefont {H.~T.}\ \bibnamefont {Cao}},
  \bibinfo {author} {\bibfnamefont {D.~B.~T.}\ \bibnamefont {Thoai}}, \ and\
  \bibinfo {author} {\bibfnamefont {H.}~\bibnamefont {Haug}},\ }\href@noop {}
  {\bibfield  {journal} {\bibinfo  {journal} {Physical Review B}\ }\textbf
  {\bibinfo {volume} {72}},\ \bibinfo {pages} {085301} (\bibinfo {year}
  {2005})}\BibitemShut {NoStop}%
\bibitem [{\citenamefont {Mazza}\ \emph {et~al.}(2013)\citenamefont {Mazza},
  \citenamefont {K{\'e}na-Cohen}, \citenamefont {Michetti},\ and\ \citenamefont
  {La~Rocca}}]{mazza2013microscopic}%
  \BibitemOpen
  \bibfield  {author} {\bibinfo {author} {\bibfnamefont {L.}~\bibnamefont
  {Mazza}}, \bibinfo {author} {\bibfnamefont {S.}~\bibnamefont
  {K{\'e}na-Cohen}}, \bibinfo {author} {\bibfnamefont {P.}~\bibnamefont
  {Michetti}}, \ and\ \bibinfo {author} {\bibfnamefont {G.~C.}\ \bibnamefont
  {La~Rocca}},\ }\href@noop {} {\bibfield  {journal} {\bibinfo  {journal}
  {Physical Review B}\ }\textbf {\bibinfo {volume} {88}},\ \bibinfo {pages}
  {075321} (\bibinfo {year} {2013})}\BibitemShut {NoStop}%
\bibitem [{\citenamefont {Kasprzak}\ \emph {et~al.}(2008)\citenamefont
  {Kasprzak}, \citenamefont {Solnyshkov}, \citenamefont {Andr{\'e}},
  \citenamefont {Dang},\ and\ \citenamefont
  {Malpuech}}]{kasprzak2008formation}%
  \BibitemOpen
  \bibfield  {author} {\bibinfo {author} {\bibfnamefont {J.}~\bibnamefont
  {Kasprzak}}, \bibinfo {author} {\bibfnamefont {D.~D.}\ \bibnamefont
  {Solnyshkov}}, \bibinfo {author} {\bibfnamefont {R.}~\bibnamefont
  {Andr{\'e}}}, \bibinfo {author} {\bibfnamefont {L.~S.}\ \bibnamefont {Dang}},
  \ and\ \bibinfo {author} {\bibfnamefont {G.}~\bibnamefont {Malpuech}},\
  }\href@noop {} {\bibfield  {journal} {\bibinfo  {journal} {Physical Review
  Letters}\ }\textbf {\bibinfo {volume} {101}},\ \bibinfo {pages} {146404}
  (\bibinfo {year} {2008})}\BibitemShut {NoStop}%
\bibitem [{\citenamefont {Chikkaraddy}\ \emph {et~al.}(2016)\citenamefont
  {Chikkaraddy}, \citenamefont {De~Nijs}, \citenamefont {Benz}, \citenamefont
  {Barrow}, \citenamefont {Scherman}, \citenamefont {Rosta}, \citenamefont
  {Demetriadou}, \citenamefont {Fox}, \citenamefont {Hess},\ and\ \citenamefont
  {Baumberg}}]{chikkaraddy2016single}%
  \BibitemOpen
  \bibfield  {author} {\bibinfo {author} {\bibfnamefont {R.}~\bibnamefont
  {Chikkaraddy}}, \bibinfo {author} {\bibfnamefont {B.}~\bibnamefont
  {De~Nijs}}, \bibinfo {author} {\bibfnamefont {F.}~\bibnamefont {Benz}},
  \bibinfo {author} {\bibfnamefont {S.~J.}\ \bibnamefont {Barrow}}, \bibinfo
  {author} {\bibfnamefont {O.~A.}\ \bibnamefont {Scherman}}, \bibinfo {author}
  {\bibfnamefont {E.}~\bibnamefont {Rosta}}, \bibinfo {author} {\bibfnamefont
  {A.}~\bibnamefont {Demetriadou}}, \bibinfo {author} {\bibfnamefont
  {P.}~\bibnamefont {Fox}}, \bibinfo {author} {\bibfnamefont {O.}~\bibnamefont
  {Hess}}, \ and\ \bibinfo {author} {\bibfnamefont {J.~J.}\ \bibnamefont
  {Baumberg}},\ }\href@noop {} {\bibfield  {journal} {\bibinfo  {journal}
  {Nature}\ }\textbf {\bibinfo {volume} {535}},\ \bibinfo {pages} {127}
  (\bibinfo {year} {2016})}\BibitemShut {NoStop}%
\bibitem [{\citenamefont {Wang}\ \emph {et~al.}(2016)\citenamefont {Wang},
  \citenamefont {Li}, \citenamefont {Chervy}, \citenamefont {Shalabney},
  \citenamefont {Azzini}, \citenamefont {Orgiu}, \citenamefont {Hutchison},
  \citenamefont {Genet}, \citenamefont {Samor{\`\i}},\ and\ \citenamefont
  {Ebbesen}}]{wang2016coherent}%
  \BibitemOpen
  \bibfield  {author} {\bibinfo {author} {\bibfnamefont {S.}~\bibnamefont
  {Wang}}, \bibinfo {author} {\bibfnamefont {S.}~\bibnamefont {Li}}, \bibinfo
  {author} {\bibfnamefont {T.}~\bibnamefont {Chervy}}, \bibinfo {author}
  {\bibfnamefont {A.}~\bibnamefont {Shalabney}}, \bibinfo {author}
  {\bibfnamefont {S.}~\bibnamefont {Azzini}}, \bibinfo {author} {\bibfnamefont
  {E.}~\bibnamefont {Orgiu}}, \bibinfo {author} {\bibfnamefont {J.~A.}\
  \bibnamefont {Hutchison}}, \bibinfo {author} {\bibfnamefont {C.}~\bibnamefont
  {Genet}}, \bibinfo {author} {\bibfnamefont {P.}~\bibnamefont {Samor{\`\i}}},
  \ and\ \bibinfo {author} {\bibfnamefont {T.~W.}\ \bibnamefont {Ebbesen}},\
  }\href@noop {} {\bibfield  {journal} {\bibinfo  {journal} {Nano Letters}\
  }\textbf {\bibinfo {volume} {16}},\ \bibinfo {pages} {4368} (\bibinfo {year}
  {2016})}\BibitemShut {NoStop}%
\bibitem [{\citenamefont {Liu}\ \emph {et~al.}(2017)\citenamefont {Liu},
  \citenamefont {Zhou}, \citenamefont {Yu}, \citenamefont {Zhang},
  \citenamefont {Wang}, \citenamefont {Liu}, \citenamefont {Wei}, \citenamefont
  {Chen},\ and\ \citenamefont {Wang}}]{liu2017strong}%
  \BibitemOpen
  \bibfield  {author} {\bibinfo {author} {\bibfnamefont {R.}~\bibnamefont
  {Liu}}, \bibinfo {author} {\bibfnamefont {Z.-K.}\ \bibnamefont {Zhou}},
  \bibinfo {author} {\bibfnamefont {Y.-C.}\ \bibnamefont {Yu}}, \bibinfo
  {author} {\bibfnamefont {T.}~\bibnamefont {Zhang}}, \bibinfo {author}
  {\bibfnamefont {H.}~\bibnamefont {Wang}}, \bibinfo {author} {\bibfnamefont
  {G.}~\bibnamefont {Liu}}, \bibinfo {author} {\bibfnamefont {Y.}~\bibnamefont
  {Wei}}, \bibinfo {author} {\bibfnamefont {H.}~\bibnamefont {Chen}}, \ and\
  \bibinfo {author} {\bibfnamefont {X.-H.}\ \bibnamefont {Wang}},\ }\href@noop
  {} {\bibfield  {journal} {\bibinfo  {journal} {Physical Review Letters}\
  }\textbf {\bibinfo {volume} {118}},\ \bibinfo {pages} {237401} (\bibinfo
  {year} {2017})}\BibitemShut {NoStop}%
\bibitem [{\citenamefont {Kleemann}\ \emph {et~al.}(2017)\citenamefont
  {Kleemann}, \citenamefont {Chikkaraddy}, \citenamefont {Alexeev},
  \citenamefont {Kos}, \citenamefont {Carnegie}, \citenamefont {Deacon},
  \citenamefont {De~Pury}, \citenamefont {Gro{\ss}e}, \citenamefont {De~Nijs},
  \citenamefont {Mertens} \emph {et~al.}}]{kleemann2017strong}%
  \BibitemOpen
  \bibfield  {author} {\bibinfo {author} {\bibfnamefont {M.-E.}\ \bibnamefont
  {Kleemann}}, \bibinfo {author} {\bibfnamefont {R.}~\bibnamefont
  {Chikkaraddy}}, \bibinfo {author} {\bibfnamefont {E.~M.}\ \bibnamefont
  {Alexeev}}, \bibinfo {author} {\bibfnamefont {D.}~\bibnamefont {Kos}},
  \bibinfo {author} {\bibfnamefont {C.}~\bibnamefont {Carnegie}}, \bibinfo
  {author} {\bibfnamefont {W.}~\bibnamefont {Deacon}}, \bibinfo {author}
  {\bibfnamefont {A.~C.}\ \bibnamefont {De~Pury}}, \bibinfo {author}
  {\bibfnamefont {C.}~\bibnamefont {Gro{\ss}e}}, \bibinfo {author}
  {\bibfnamefont {B.}~\bibnamefont {De~Nijs}}, \bibinfo {author} {\bibfnamefont
  {J.}~\bibnamefont {Mertens}},  \emph {et~al.},\ }\href@noop {} {\bibfield
  {journal} {\bibinfo  {journal} {Nature Communications}\ }\textbf {\bibinfo
  {volume} {8}},\ \bibinfo {pages} {1} (\bibinfo {year} {2017})}\BibitemShut
  {NoStop}%
\bibitem [{\citenamefont {St\"uhrenberg}\ \emph {et~al.}(2018)\citenamefont
  {St\"uhrenberg}, \citenamefont {Munkhbat}, \citenamefont {Baranov},
  \citenamefont {Cuadra}, \citenamefont {Yankovich}, \citenamefont
  {Antosiewicz}, \citenamefont {Olsson},\ and\ \citenamefont
  {Shegai}}]{stuhrenberg2018strong}%
  \BibitemOpen
  \bibfield  {author} {\bibinfo {author} {\bibfnamefont {M.}~\bibnamefont
  {St\"uhrenberg}}, \bibinfo {author} {\bibfnamefont {B.}~\bibnamefont
  {Munkhbat}}, \bibinfo {author} {\bibfnamefont {D.~G.}\ \bibnamefont
  {Baranov}}, \bibinfo {author} {\bibfnamefont {J.}~\bibnamefont {Cuadra}},
  \bibinfo {author} {\bibfnamefont {A.~B.}\ \bibnamefont {Yankovich}}, \bibinfo
  {author} {\bibfnamefont {T.~J.}\ \bibnamefont {Antosiewicz}}, \bibinfo
  {author} {\bibfnamefont {E.}~\bibnamefont {Olsson}}, \ and\ \bibinfo {author}
  {\bibfnamefont {T.}~\bibnamefont {Shegai}},\ }\href@noop {} {\bibfield
  {journal} {\bibinfo  {journal} {Nano Letters}\ }\textbf {\bibinfo {volume}
  {18}},\ \bibinfo {pages} {5938} (\bibinfo {year} {2018})}\BibitemShut
  {NoStop}%
\bibitem [{\citenamefont {Han}\ \emph {et~al.}(2018)\citenamefont {Han},
  \citenamefont {Wang}, \citenamefont {Xing}, \citenamefont {Wang},\ and\
  \citenamefont {Lu}}]{han2018rabi}%
  \BibitemOpen
  \bibfield  {author} {\bibinfo {author} {\bibfnamefont {X.}~\bibnamefont
  {Han}}, \bibinfo {author} {\bibfnamefont {K.}~\bibnamefont {Wang}}, \bibinfo
  {author} {\bibfnamefont {X.}~\bibnamefont {Xing}}, \bibinfo {author}
  {\bibfnamefont {M.}~\bibnamefont {Wang}}, \ and\ \bibinfo {author}
  {\bibfnamefont {P.}~\bibnamefont {Lu}},\ }\href@noop {} {\bibfield  {journal}
  {\bibinfo  {journal} {ACS Photonics}\ }\textbf {\bibinfo {volume} {5}},\
  \bibinfo {pages} {3970} (\bibinfo {year} {2018})}\BibitemShut {NoStop}%
\bibitem [{\citenamefont {Geisler}\ \emph {et~al.}(2019)\citenamefont
  {Geisler}, \citenamefont {Cui}, \citenamefont {Wang}, \citenamefont
  {Rindzevicius}, \citenamefont {Gammelgaard}, \citenamefont {Jessen},
  \citenamefont {Goncalves}, \citenamefont {Todisco}, \citenamefont
  {B{\o}ggild}, \citenamefont {Boisen}, \citenamefont {Wubs}, \citenamefont
  {Mortensen}, \citenamefont {Xiao},\ and\ \citenamefont
  {Stenger}}]{geisler2019single}%
  \BibitemOpen
  \bibfield  {author} {\bibinfo {author} {\bibfnamefont {M.}~\bibnamefont
  {Geisler}}, \bibinfo {author} {\bibfnamefont {X.}~\bibnamefont {Cui}},
  \bibinfo {author} {\bibfnamefont {J.}~\bibnamefont {Wang}}, \bibinfo {author}
  {\bibfnamefont {T.}~\bibnamefont {Rindzevicius}}, \bibinfo {author}
  {\bibfnamefont {L.}~\bibnamefont {Gammelgaard}}, \bibinfo {author}
  {\bibfnamefont {B.~S.}\ \bibnamefont {Jessen}}, \bibinfo {author}
  {\bibfnamefont {P.~A.~D.}\ \bibnamefont {Goncalves}}, \bibinfo {author}
  {\bibfnamefont {F.}~\bibnamefont {Todisco}}, \bibinfo {author} {\bibfnamefont
  {P.}~\bibnamefont {B{\o}ggild}}, \bibinfo {author} {\bibfnamefont
  {A.}~\bibnamefont {Boisen}}, \bibinfo {author} {\bibfnamefont
  {M.}~\bibnamefont {Wubs}}, \bibinfo {author} {\bibfnamefont {N.~A.}\
  \bibnamefont {Mortensen}}, \bibinfo {author} {\bibfnamefont {S.}~\bibnamefont
  {Xiao}}, \ and\ \bibinfo {author} {\bibfnamefont {N.}~\bibnamefont
  {Stenger}},\ }\href@noop {} {\bibfield  {journal} {\bibinfo  {journal} {{ACS
  Photonics}}\ }\textbf {\bibinfo {volume} {6}},\ \bibinfo {pages} {994}
  (\bibinfo {year} {2019})}\BibitemShut {NoStop}%
\bibitem [{\citenamefont {Qin}\ \emph {et~al.}(2020)\citenamefont {Qin},
  \citenamefont {Chen}, \citenamefont {Zhang}, \citenamefont {Zhang},
  \citenamefont {Blaikie}, \citenamefont {Ding},\ and\ \citenamefont
  {Qiu}}]{qin2020revealing}%
  \BibitemOpen
  \bibfield  {author} {\bibinfo {author} {\bibfnamefont {J.}~\bibnamefont
  {Qin}}, \bibinfo {author} {\bibfnamefont {Y.-H.}\ \bibnamefont {Chen}},
  \bibinfo {author} {\bibfnamefont {Z.}~\bibnamefont {Zhang}}, \bibinfo
  {author} {\bibfnamefont {Y.}~\bibnamefont {Zhang}}, \bibinfo {author}
  {\bibfnamefont {R.~J.}\ \bibnamefont {Blaikie}}, \bibinfo {author}
  {\bibfnamefont {B.}~\bibnamefont {Ding}}, \ and\ \bibinfo {author}
  {\bibfnamefont {M.}~\bibnamefont {Qiu}},\ }\href@noop {} {\bibfield
  {journal} {\bibinfo  {journal} {Physical Review Letters}\ }\textbf {\bibinfo
  {volume} {124}},\ \bibinfo {pages} {063902} (\bibinfo {year}
  {2020})}\BibitemShut {NoStop}%
\bibitem [{\citenamefont {Gro{\ss}}\ \emph {et~al.}(2018)\citenamefont
  {Gro{\ss}}, \citenamefont {Hamm}, \citenamefont {Tufarelli}, \citenamefont
  {Hess},\ and\ \citenamefont {Hecht}}]{gross2018near}%
  \BibitemOpen
  \bibfield  {author} {\bibinfo {author} {\bibfnamefont {H.}~\bibnamefont
  {Gro{\ss}}}, \bibinfo {author} {\bibfnamefont {J.~M.}\ \bibnamefont {Hamm}},
  \bibinfo {author} {\bibfnamefont {T.}~\bibnamefont {Tufarelli}}, \bibinfo
  {author} {\bibfnamefont {O.}~\bibnamefont {Hess}}, \ and\ \bibinfo {author}
  {\bibfnamefont {B.}~\bibnamefont {Hecht}},\ }\href@noop {} {\bibfield
  {journal} {\bibinfo  {journal} {Science advances}\ }\textbf {\bibinfo
  {volume} {4}} (\bibinfo {year} {2018})}\BibitemShut {NoStop}%
\bibitem [{\citenamefont {Choi}\ \emph {et~al.}(2017)\citenamefont {Choi},
  \citenamefont {Heuck},\ and\ \citenamefont
  {Englund}}]{self_similar_Choi_2017}%
  \BibitemOpen
  \bibfield  {author} {\bibinfo {author} {\bibfnamefont {H.}~\bibnamefont
  {Choi}}, \bibinfo {author} {\bibfnamefont {M.}~\bibnamefont {Heuck}}, \ and\
  \bibinfo {author} {\bibfnamefont {D.}~\bibnamefont {Englund}},\ }\href
  {\doibase 10.1103/PhysRevLett.118.223605} {\bibfield  {journal} {\bibinfo
  {journal} {Physical Review Letters}\ }\textbf {\bibinfo {volume} {118}},\
  \bibinfo {pages} {223605} (\bibinfo {year} {2017})}\BibitemShut {NoStop}%
\bibitem [{\citenamefont {Hu}\ \emph {et~al.}(2018)\citenamefont {Hu},
  \citenamefont {Khater}, \citenamefont {Salas-Montiel}, \citenamefont
  {Kratschmer}, \citenamefont {Engelmann}, \citenamefont {Green},\ and\
  \citenamefont {Weiss}}]{Shuren_exp_sub_2018}%
  \BibitemOpen
  \bibfield  {author} {\bibinfo {author} {\bibfnamefont {S.}~\bibnamefont
  {Hu}}, \bibinfo {author} {\bibfnamefont {M.}~\bibnamefont {Khater}}, \bibinfo
  {author} {\bibfnamefont {R.}~\bibnamefont {Salas-Montiel}}, \bibinfo {author}
  {\bibfnamefont {E.}~\bibnamefont {Kratschmer}}, \bibinfo {author}
  {\bibfnamefont {S.}~\bibnamefont {Engelmann}}, \bibinfo {author}
  {\bibfnamefont {W.~M.~J.}\ \bibnamefont {Green}}, \ and\ \bibinfo {author}
  {\bibfnamefont {S.~M.}\ \bibnamefont {Weiss}},\ }\href@noop {} {\bibfield
  {journal} {\bibinfo  {journal} {Science Advances}\ }\textbf {\bibinfo
  {volume} {4}} (\bibinfo {year} {2018})}\BibitemShut {NoStop}%
\bibitem [{\citenamefont {Wang}\ \emph {et~al.}(2018)\citenamefont {Wang},
  \citenamefont {Christiansen}, \citenamefont {Yu}, \citenamefont {Mørk},\
  and\ \citenamefont {Sigmund}}]{jesm_max_qual}%
  \BibitemOpen
  \bibfield  {author} {\bibinfo {author} {\bibfnamefont {F.}~\bibnamefont
  {Wang}}, \bibinfo {author} {\bibfnamefont {R.~E.}\ \bibnamefont
  {Christiansen}}, \bibinfo {author} {\bibfnamefont {Y.}~\bibnamefont {Yu}},
  \bibinfo {author} {\bibfnamefont {J.}~\bibnamefont {Mørk}}, \ and\ \bibinfo
  {author} {\bibfnamefont {O.}~\bibnamefont {Sigmund}},\ }\href@noop {}
  {\bibfield  {journal} {\bibinfo  {journal} {Applied Physics Letters}\
  }\textbf {\bibinfo {volume} {113}},\ \bibinfo {pages} {241101} (\bibinfo
  {year} {2018})}\BibitemShut {NoStop}%
\bibitem [{\citenamefont {Hornecker}\ \emph
  {et~al.}(2017{\natexlab{a}})\citenamefont {Hornecker}, \citenamefont
  {Auff\`eves},\ and\ \citenamefont {Grange}}]{tens_prev_1}%
  \BibitemOpen
  \bibfield  {author} {\bibinfo {author} {\bibfnamefont {G.}~\bibnamefont
  {Hornecker}}, \bibinfo {author} {\bibfnamefont {A.}~\bibnamefont
  {Auff\`eves}}, \ and\ \bibinfo {author} {\bibfnamefont {T.}~\bibnamefont
  {Grange}},\ }\href {\doibase 10.1103/PhysRevB.95.035404} {\bibfield
  {journal} {\bibinfo  {journal} {Phys. Rev. B}\ }\textbf {\bibinfo {volume}
  {95}},\ \bibinfo {pages} {035404} (\bibinfo {year}
  {2017}{\natexlab{a}})}\BibitemShut {NoStop}%
\bibitem [{\citenamefont {Morreau}\ and\ \citenamefont
  {Muljarov}(2019)}]{tens_prev_2}%
  \BibitemOpen
  \bibfield  {author} {\bibinfo {author} {\bibfnamefont {A.}~\bibnamefont
  {Morreau}}\ and\ \bibinfo {author} {\bibfnamefont {E.~A.}\ \bibnamefont
  {Muljarov}},\ }\href@noop {} {\bibfield  {journal} {\bibinfo  {journal}
  {Phys. Rev. B}\ }\textbf {\bibinfo {volume} {100}},\ \bibinfo {pages}
  {115309} (\bibinfo {year} {2019})}\BibitemShut {NoStop}%
\bibitem [{\citenamefont {Vagov}\ \emph {et~al.}(2011)\citenamefont {Vagov},
  \citenamefont {Croitoru}, \citenamefont {Gl\"assl}, \citenamefont {Axt},\
  and\ \citenamefont {Kuhn}}]{tens_prev_3}%
  \BibitemOpen
  \bibfield  {author} {\bibinfo {author} {\bibfnamefont {A.}~\bibnamefont
  {Vagov}}, \bibinfo {author} {\bibfnamefont {M.~D.}\ \bibnamefont {Croitoru}},
  \bibinfo {author} {\bibfnamefont {M.}~\bibnamefont {Gl\"assl}}, \bibinfo
  {author} {\bibfnamefont {V.~M.}\ \bibnamefont {Axt}}, \ and\ \bibinfo
  {author} {\bibfnamefont {T.}~\bibnamefont {Kuhn}},\ }\href {\doibase
  10.1103/PhysRevB.83.094303} {\bibfield  {journal} {\bibinfo  {journal} {Phys.
  Rev. B}\ }\textbf {\bibinfo {volume} {83}},\ \bibinfo {pages} {094303}
  (\bibinfo {year} {2011})}\BibitemShut {NoStop}%
\bibitem [{\citenamefont {Kaer}\ \emph {et~al.}(2010)\citenamefont {Kaer},
  \citenamefont {Nielsen}, \citenamefont {Lodahl}, \citenamefont {Jauho},\ and\
  \citenamefont {M\o{}rk}}]{tens_prev_4}%
  \BibitemOpen
  \bibfield  {author} {\bibinfo {author} {\bibfnamefont {P.}~\bibnamefont
  {Kaer}}, \bibinfo {author} {\bibfnamefont {T.~R.}\ \bibnamefont {Nielsen}},
  \bibinfo {author} {\bibfnamefont {P.}~\bibnamefont {Lodahl}}, \bibinfo
  {author} {\bibfnamefont {A.-P.}\ \bibnamefont {Jauho}}, \ and\ \bibinfo
  {author} {\bibfnamefont {J.}~\bibnamefont {M\o{}rk}},\ }\href {\doibase
  10.1103/PhysRevLett.104.157401} {\bibfield  {journal} {\bibinfo  {journal}
  {Physical Review Letters}\ }\textbf {\bibinfo {volume} {104}},\ \bibinfo
  {pages} {157401} (\bibinfo {year} {2010})}\BibitemShut {NoStop}%
\bibitem [{\citenamefont {Kaer}\ \emph {et~al.}(2013)\citenamefont {Kaer},
  \citenamefont {Lodahl}, \citenamefont {Jauho},\ and\ \citenamefont
  {Mork}}]{tens_prev_5}%
  \BibitemOpen
  \bibfield  {author} {\bibinfo {author} {\bibfnamefont {P.}~\bibnamefont
  {Kaer}}, \bibinfo {author} {\bibfnamefont {P.}~\bibnamefont {Lodahl}},
  \bibinfo {author} {\bibfnamefont {A.-P.}\ \bibnamefont {Jauho}}, \ and\
  \bibinfo {author} {\bibfnamefont {J.}~\bibnamefont {Mork}},\ }\href {\doibase
  10.1103/PhysRevB.87.081308} {\bibfield  {journal} {\bibinfo  {journal} {Phys.
  Rev. B}\ }\textbf {\bibinfo {volume} {87}},\ \bibinfo {pages} {081308(R)}
  (\bibinfo {year} {2013})}\BibitemShut {NoStop}%
\bibitem [{\citenamefont {J\o{}rgensen}\ and\ \citenamefont
  {Pollock}(2019)}]{Emil_tensor1}%
  \BibitemOpen
  \bibfield  {author} {\bibinfo {author} {\bibfnamefont {M.~R.}\ \bibnamefont
  {J\o{}rgensen}}\ and\ \bibinfo {author} {\bibfnamefont {F.~A.}\ \bibnamefont
  {Pollock}},\ }\href {\doibase 10.1103/PhysRevLett.123.240602} {\bibfield
  {journal} {\bibinfo  {journal} {Physical Review Letters}\ }\textbf {\bibinfo
  {volume} {123}},\ \bibinfo {pages} {240602} (\bibinfo {year}
  {2019})}\BibitemShut {NoStop}%
\bibitem [{\citenamefont {Strathearn}\ \emph {et~al.}(2018)\citenamefont
  {Strathearn}, \citenamefont {Kirton}, \citenamefont {Kilda}, \citenamefont
  {Keeling},\ and\ \citenamefont {Lovett}}]{Emil_tensor2}%
  \BibitemOpen
  \bibfield  {author} {\bibinfo {author} {\bibfnamefont {A.}~\bibnamefont
  {Strathearn}}, \bibinfo {author} {\bibfnamefont {P.}~\bibnamefont {Kirton}},
  \bibinfo {author} {\bibfnamefont {D.}~\bibnamefont {Kilda}}, \bibinfo
  {author} {\bibfnamefont {J.}~\bibnamefont {Keeling}}, \ and\ \bibinfo
  {author} {\bibfnamefont {B.~W.}\ \bibnamefont {Lovett}},\ }\href@noop {}
  {\bibfield  {journal} {\bibinfo  {journal} {Nature Communications}\ }\textbf
  {\bibinfo {volume} {9}} (\bibinfo {year} {2018})}\BibitemShut {NoStop}%
\bibitem [{\citenamefont {Wilson-Rae}\ and\ \citenamefont
  {Imamo\ifmmode~\breve{g}\else \u{g}\fi{}lu}(2002)}]{phonon_1}%
  \BibitemOpen
  \bibfield  {author} {\bibinfo {author} {\bibfnamefont {I.}~\bibnamefont
  {Wilson-Rae}}\ and\ \bibinfo {author} {\bibfnamefont {A.}~\bibnamefont
  {Imamo\ifmmode~\breve{g}\else \u{g}\fi{}lu}},\ }\href {\doibase
  10.1103/PhysRevB.65.235311} {\bibfield  {journal} {\bibinfo  {journal} {Phys.
  Rev. B}\ }\textbf {\bibinfo {volume} {65}},\ \bibinfo {pages} {235311}
  (\bibinfo {year} {2002})}\BibitemShut {NoStop}%
\bibitem [{\citenamefont {Laucht}\ \emph {et~al.}(2011)\citenamefont {Laucht},
  \citenamefont {Hauke}, \citenamefont {Neumann}, \citenamefont {Günthner},
  \citenamefont {Hofbauer}, \citenamefont {Mohtashami}, \citenamefont
  {Müller}, \citenamefont {Böhm}, \citenamefont {Bichler}, \citenamefont
  {Amann}, \citenamefont {Kaniber},\ and\ \citenamefont {Finley}}]{phonon_2}%
  \BibitemOpen
  \bibfield  {author} {\bibinfo {author} {\bibfnamefont {A.}~\bibnamefont
  {Laucht}}, \bibinfo {author} {\bibfnamefont {N.}~\bibnamefont {Hauke}},
  \bibinfo {author} {\bibfnamefont {A.}~\bibnamefont {Neumann}}, \bibinfo
  {author} {\bibfnamefont {T.}~\bibnamefont {Günthner}}, \bibinfo {author}
  {\bibfnamefont {F.}~\bibnamefont {Hofbauer}}, \bibinfo {author}
  {\bibfnamefont {A.}~\bibnamefont {Mohtashami}}, \bibinfo {author}
  {\bibfnamefont {K.}~\bibnamefont {Müller}}, \bibinfo {author} {\bibfnamefont
  {G.}~\bibnamefont {Böhm}}, \bibinfo {author} {\bibfnamefont
  {M.}~\bibnamefont {Bichler}}, \bibinfo {author} {\bibfnamefont {M.-C.}\
  \bibnamefont {Amann}}, \bibinfo {author} {\bibfnamefont {M.}~\bibnamefont
  {Kaniber}}, \ and\ \bibinfo {author} {\bibfnamefont {J.~J.}\ \bibnamefont
  {Finley}},\ }\href@noop {} {\bibfield  {journal} {\bibinfo  {journal}
  {Journal of Applied Physics}\ }\textbf {\bibinfo {volume} {109}},\ \bibinfo
  {pages} {102404} (\bibinfo {year} {2011})}\BibitemShut {NoStop}%
\bibitem [{\citenamefont {Besombes}\ \emph
  {et~al.}(2001{\natexlab{b}})\citenamefont {Besombes}, \citenamefont {Kheng},
  \citenamefont {Marsal},\ and\ \citenamefont {Mariette}}]{phonon_3}%
  \BibitemOpen
  \bibfield  {author} {\bibinfo {author} {\bibfnamefont {L.}~\bibnamefont
  {Besombes}}, \bibinfo {author} {\bibfnamefont {K.}~\bibnamefont {Kheng}},
  \bibinfo {author} {\bibfnamefont {L.}~\bibnamefont {Marsal}}, \ and\ \bibinfo
  {author} {\bibfnamefont {H.}~\bibnamefont {Mariette}},\ }\href {\doibase
  10.1103/PhysRevB.63.155307} {\bibfield  {journal} {\bibinfo  {journal} {Phys.
  Rev. B}\ }\textbf {\bibinfo {volume} {63}},\ \bibinfo {pages} {155307}
  (\bibinfo {year} {2001}{\natexlab{b}})}\BibitemShut {NoStop}%
\bibitem [{\citenamefont {Senellart}\ \emph {et~al.}(2017)\citenamefont
  {Senellart}, \citenamefont {Solomon},\ and\ \citenamefont
  {White}}]{phonon_4}%
  \BibitemOpen
  \bibfield  {author} {\bibinfo {author} {\bibfnamefont {P.}~\bibnamefont
  {Senellart}}, \bibinfo {author} {\bibfnamefont {G.}~\bibnamefont {Solomon}},
  \ and\ \bibinfo {author} {\bibfnamefont {A.}~\bibnamefont {White}},\ }\href
  {https://doi.org/10.1038/nnano.2017.218} {\bibfield  {journal} {\bibinfo
  {journal} {Nature nanotechnology}\ }\textbf {\bibinfo {volume} {12}},\
  \bibinfo {pages} {1026} (\bibinfo {year} {2017})}\BibitemShut {NoStop}%
\bibitem [{\citenamefont {McCutcheon}(2016)}]{dara_2time_sens}%
  \BibitemOpen
  \bibfield  {author} {\bibinfo {author} {\bibfnamefont {D.~P.~S.}\
  \bibnamefont {McCutcheon}},\ }\href {\doibase 10.1103/PhysRevA.93.022119}
  {\bibfield  {journal} {\bibinfo  {journal} {Phys. Rev. A}\ }\textbf {\bibinfo
  {volume} {93}},\ \bibinfo {pages} {022119} (\bibinfo {year}
  {2016})}\BibitemShut {NoStop}%
\bibitem [{\citenamefont {Pollock}\ \emph {et~al.}(2018)\citenamefont
  {Pollock}, \citenamefont {Rodr{\'\i}guez-Rosario}, \citenamefont
  {Frauenheim}, \citenamefont {Paternostro},\ and\ \citenamefont
  {Modi}}]{pollock2018operational}%
  \BibitemOpen
  \bibfield  {author} {\bibinfo {author} {\bibfnamefont {F.~A.}\ \bibnamefont
  {Pollock}}, \bibinfo {author} {\bibfnamefont {C.}~\bibnamefont
  {Rodr{\'\i}guez-Rosario}}, \bibinfo {author} {\bibfnamefont {T.}~\bibnamefont
  {Frauenheim}}, \bibinfo {author} {\bibfnamefont {M.}~\bibnamefont
  {Paternostro}}, \ and\ \bibinfo {author} {\bibfnamefont {K.}~\bibnamefont
  {Modi}},\ }\href@noop {} {\bibfield  {journal} {\bibinfo  {journal} {Physical
  Review Letters}\ }\textbf {\bibinfo {volume} {120}},\ \bibinfo {pages}
  {040405} (\bibinfo {year} {2018})}\BibitemShut {NoStop}%
\bibitem [{\citenamefont {Denning}\ \emph
  {et~al.}(2020{\natexlab{b}})\citenamefont {Denning}, \citenamefont
  {Iles-Smith}, \citenamefont {Gregersen},\ and\ \citenamefont
  {Mork}}]{phononeffects_emil}%
  \BibitemOpen
  \bibfield  {author} {\bibinfo {author} {\bibfnamefont {E.}~\bibnamefont
  {Denning}}, \bibinfo {author} {\bibfnamefont {J.}~\bibnamefont {Iles-Smith}},
  \bibinfo {author} {\bibfnamefont {N.}~\bibnamefont {Gregersen}}, \ and\
  \bibinfo {author} {\bibfnamefont {J.}~\bibnamefont {Mork}},\ }\href {\doibase
  10.1364/OME.380601} {\bibfield  {journal} {\bibinfo  {journal} {Optical
  Materials Express}\ }\textbf {\bibinfo {volume} {10}},\ \bibinfo {pages}
  {222} (\bibinfo {year} {2020}{\natexlab{b}})}\BibitemShut {NoStop}%
\bibitem [{\citenamefont {Mahan}(2000)}]{manyparticle}%
  \BibitemOpen
  \bibfield  {author} {\bibinfo {author} {\bibfnamefont {G.}~\bibnamefont
  {Mahan}},\ }\enquote {\bibinfo {title} {Many-particle physics},}\ \ (\bibinfo
   {publisher} {Springer Science and Business Media},\ \bibinfo {year} {2000})\
  pp.\ \bibinfo {pages} {218--222}\BibitemShut {NoStop}%
\bibitem [{\citenamefont {Iles-Smith}\ \emph
  {et~al.}(2017{\natexlab{b}})\citenamefont {Iles-Smith}, \citenamefont
  {McCutcheon}, \citenamefont {Mørk},\ and\ \citenamefont
  {Nazir}}]{Iles_Smith_2017_Limits_to_coherent}%
  \BibitemOpen
  \bibfield  {author} {\bibinfo {author} {\bibfnamefont {J.}~\bibnamefont
  {Iles-Smith}}, \bibinfo {author} {\bibfnamefont {D.~P.~S.}\ \bibnamefont
  {McCutcheon}}, \bibinfo {author} {\bibfnamefont {J.}~\bibnamefont {M{\o}rk}}, \
  and\ \bibinfo {author} {\bibfnamefont {A.}~\bibnamefont {Nazir}},\
  }\href@noop {} {\bibfield  {journal} {\bibinfo  {journal} {Physical Review
  B}\ }\textbf {\bibinfo {volume} {95}},\ \bibinfo {pages} {201305(R)}
  (\bibinfo {year} {2017}{\natexlab{b}})}\BibitemShut {NoStop}%
\bibitem [{\citenamefont {Kaer}\ \emph {et~al.}(2012)\citenamefont {Kaer},
  \citenamefont {Nielsen}, \citenamefont {Lodahl}, \citenamefont {Jauho},\ and\
  \citenamefont {M{\o}rk}}]{kaer2012microscopic}%
  \BibitemOpen
  \bibfield  {author} {\bibinfo {author} {\bibfnamefont {P.}~\bibnamefont
  {Kaer}}, \bibinfo {author} {\bibfnamefont {T.~R.}\ \bibnamefont {Nielsen}},
  \bibinfo {author} {\bibfnamefont {P.}~\bibnamefont {Lodahl}}, \bibinfo
  {author} {\bibfnamefont {A.-P.}\ \bibnamefont {Jauho}}, \ and\ \bibinfo
  {author} {\bibfnamefont {J.}~\bibnamefont {M{\o}rk}},\ }\href@noop {}
  {\bibfield  {journal} {\bibinfo  {journal} {Physical Review B}\ }\textbf
  {\bibinfo {volume} {86}},\ \bibinfo {pages} {085302} (\bibinfo {year}
  {2012})}\BibitemShut {NoStop}%
\bibitem [{\citenamefont {W{\"u}rger}(1998)}]{wurger1998strong}%
  \BibitemOpen
  \bibfield  {author} {\bibinfo {author} {\bibfnamefont {A.}~\bibnamefont
  {W{\"u}rger}},\ }\href@noop {} {\bibfield  {journal} {\bibinfo  {journal}
  {Physical Review B}\ }\textbf {\bibinfo {volume} {57}},\ \bibinfo {pages}
  {347} (\bibinfo {year} {1998})}\BibitemShut {NoStop}%
\bibitem [{\citenamefont {Duke}\ and\ \citenamefont
  {Mahan}(1965)}]{duke1965phonon}%
  \BibitemOpen
  \bibfield  {author} {\bibinfo {author} {\bibfnamefont {C.}~\bibnamefont
  {Duke}}\ and\ \bibinfo {author} {\bibfnamefont {G.}~\bibnamefont {Mahan}},\
  }\href@noop {} {\bibfield  {journal} {\bibinfo  {journal} {Physical Review}\
  }\textbf {\bibinfo {volume} {139}},\ \bibinfo {pages} {A1965} (\bibinfo
  {year} {1965})}\BibitemShut {NoStop}%
\bibitem [{\citenamefont {Merrifield}(1964)}]{merrifield1964theory}%
  \BibitemOpen
  \bibfield  {author} {\bibinfo {author} {\bibfnamefont {R.}~\bibnamefont
  {Merrifield}},\ }\href@noop {} {\bibfield  {journal} {\bibinfo  {journal}
  {The Journal of Chemical Physics}\ }\textbf {\bibinfo {volume} {40}},\
  \bibinfo {pages} {445} (\bibinfo {year} {1964})}\BibitemShut {NoStop}%
\bibitem [{\citenamefont {Harris}\ and\ \citenamefont
  {Silbey}(1985)}]{Silbey_variational}%
  \BibitemOpen
  \bibfield  {author} {\bibinfo {author} {\bibfnamefont {R.~A.}\ \bibnamefont
  {Harris}}\ and\ \bibinfo {author} {\bibfnamefont {R.}~\bibnamefont
  {Silbey}},\ }\href@noop {} {\bibfield  {journal} {\bibinfo  {journal} {The
  Journal of Chemical Physics}\ }\textbf {\bibinfo {volume} {83}},\ \bibinfo
  {pages} {1069} (\bibinfo {year} {1985})}\BibitemShut {NoStop}%
\bibitem [{\citenamefont {G\'omez-S\'anchez}\ and\ \citenamefont
  {Ram\'{\i}rez}(2018)}]{gomez_2018_variational_master_equation}%
  \BibitemOpen
  \bibfield  {author} {\bibinfo {author} {\bibfnamefont {O.~J.}\ \bibnamefont
  {G\'omez-S\'anchez}}\ and\ \bibinfo {author} {\bibfnamefont {H.~Y.}\
  \bibnamefont {Ram\'{\i}rez}},\ }\href {\doibase 10.1103/PhysRevA.98.053846}
  {\bibfield  {journal} {\bibinfo  {journal} {Phys. Rev. A}\ }\textbf {\bibinfo
  {volume} {98}},\ \bibinfo {pages} {053846} (\bibinfo {year}
  {2018})}\BibitemShut {NoStop}%
\bibitem [{\citenamefont {McCutcheon}\ \emph {et~al.}(2011)\citenamefont
  {McCutcheon}, \citenamefont {Dattani}, \citenamefont {Gauger}, \citenamefont
  {Lovett},\ and\ \citenamefont {Nazir}}]{mccutcheon2011general}%
  \BibitemOpen
  \bibfield  {author} {\bibinfo {author} {\bibfnamefont {D.~P.~S.}\
  \bibnamefont {McCutcheon}}, \bibinfo {author} {\bibfnamefont {N.~S.}\
  \bibnamefont {Dattani}}, \bibinfo {author} {\bibfnamefont {E.~M.}\
  \bibnamefont {Gauger}}, \bibinfo {author} {\bibfnamefont {B.~W.}\
  \bibnamefont {Lovett}}, \ and\ \bibinfo {author} {\bibfnamefont
  {A.}~\bibnamefont {Nazir}},\ }\href {\doibase 10.1103/PhysRevB.84.081305}
  {\bibfield  {journal} {\bibinfo  {journal} {Phys. Rev. B}\ }\textbf {\bibinfo
  {volume} {84}},\ \bibinfo {pages} {081305(R)} (\bibinfo {year}
  {2011})}\BibitemShut {NoStop}%
\bibitem [{\citenamefont {Nazir}\ and\ \citenamefont
  {McCutcheon}(2016)}]{nazir2016modelling}%
  \BibitemOpen
  \bibfield  {author} {\bibinfo {author} {\bibfnamefont {A.}~\bibnamefont
  {Nazir}}\ and\ \bibinfo {author} {\bibfnamefont {D.}~\bibnamefont
  {McCutcheon}},\ }\href@noop {} {\bibfield  {journal} {\bibinfo  {journal}
  {Journal of Physics: Condensed Matter}\ }\textbf {\bibinfo {volume} {28}}
  (\bibinfo {year} {2016})}\BibitemShut {NoStop}%
\bibitem [{\citenamefont {Breuer}\ \emph {et~al.}(2002)\citenamefont {Breuer},
  \citenamefont {Petruccione} \emph {et~al.}}]{breuer2002theory}%
  \BibitemOpen
  \bibfield  {author} {\bibinfo {author} {\bibfnamefont {H.-P.}\ \bibnamefont
  {Breuer}}, \bibinfo {author} {\bibfnamefont {F.}~\bibnamefont {Petruccione}},
   \emph {et~al.},\ }\href@noop {} {\emph {\bibinfo {title} {The theory of open
  quantum systems}}}\ (\bibinfo  {publisher} {Oxford University Press on
  Demand},\ \bibinfo {year} {2002})\BibitemShut {NoStop}%
\bibitem [{\citenamefont {Reigue}\ \emph {et~al.}(2017)\citenamefont {Reigue},
  \citenamefont {Iles-Smith}, \citenamefont {Lux}, \citenamefont {Monniello},
  \citenamefont {Bernard}, \citenamefont {Margaillan}, \citenamefont
  {Lemaitre}, \citenamefont {Martinez}, \citenamefont {McCutcheon},
  \citenamefont {M{\o}rk}, \citenamefont {Hostein},\ and\ \citenamefont
  {Voliotis}}]{Phonon_dephasing_Reigue}%
  \BibitemOpen
  \bibfield  {author} {\bibinfo {author} {\bibfnamefont {A.}~\bibnamefont
  {Reigue}}, \bibinfo {author} {\bibfnamefont {J.}~\bibnamefont {Iles-Smith}},
  \bibinfo {author} {\bibfnamefont {F.}~\bibnamefont {Lux}}, \bibinfo {author}
  {\bibfnamefont {L.}~\bibnamefont {Monniello}}, \bibinfo {author}
  {\bibfnamefont {M.}~\bibnamefont {Bernard}}, \bibinfo {author} {\bibfnamefont
  {F.}~\bibnamefont {Margaillan}}, \bibinfo {author} {\bibfnamefont
  {A.}~\bibnamefont {Lemaitre}}, \bibinfo {author} {\bibfnamefont
  {A.}~\bibnamefont {Martinez}}, \bibinfo {author} {\bibfnamefont
  {D. P. S.}~\bibnamefont {McCutcheon}}, \bibinfo {author} {\bibfnamefont
  {J.}~\bibnamefont {M{\o}rk}}, \bibinfo {author} {\bibfnamefont
  {R.}~\bibnamefont {Hostein}}, \ and\ \bibinfo {author} {\bibfnamefont
  {V.}~\bibnamefont {Voliotis}},\ }\href@noop {} {\bibfield  {journal}
  {\bibinfo  {journal} {Physical Review Letters}\ }\textbf {\bibinfo {volume}
  {118}},\ \bibinfo {pages} {233602} (\bibinfo {year} {2017})}\BibitemShut
  {NoStop}%
\bibitem [{\citenamefont {Tighineanu}\ \emph {et~al.}(2018)\citenamefont
  {Tighineanu}, \citenamefont {Dree\ss{}en}, \citenamefont {Flindt},
  \citenamefont {Lodahl},\ and\ \citenamefont
  {S\o{}rensen}}]{Phonon_dephasing_Tighineanu}%
  \BibitemOpen
  \bibfield  {author} {\bibinfo {author} {\bibfnamefont {P.}~\bibnamefont
  {Tighineanu}}, \bibinfo {author} {\bibfnamefont {C.~L.}\ \bibnamefont
  {Dree\ss{}en}}, \bibinfo {author} {\bibfnamefont {C.}~\bibnamefont {Flindt}},
  \bibinfo {author} {\bibfnamefont {P.}~\bibnamefont {Lodahl}}, \ and\ \bibinfo
  {author} {\bibfnamefont {A.~S.}\ \bibnamefont {S\o{}rensen}},\ }\href
  {\doibase 10.1103/PhysRevLett.120.257401} {\bibfield  {journal} {\bibinfo
  {journal} {Physical Review Letters}\ }\textbf {\bibinfo {volume} {120}},\
  \bibinfo {pages} {257401} (\bibinfo {year} {2018})}\BibitemShut {NoStop}%
\bibitem [{\citenamefont {Muljarov}\ and\ \citenamefont
  {Zimmermann}(2004)}]{Phonon_dephasing_Muljarov}%
  \BibitemOpen
  \bibfield  {author} {\bibinfo {author} {\bibfnamefont {E.~A.}\ \bibnamefont
  {Muljarov}}\ and\ \bibinfo {author} {\bibfnamefont {R.}~\bibnamefont
  {Zimmermann}},\ }\href {\doibase 10.1103/PhysRevLett.93.237401} {\bibfield
  {journal} {\bibinfo  {journal} {Physical Review Letters}\ }\textbf {\bibinfo
  {volume} {93}},\ \bibinfo {pages} {237401} (\bibinfo {year}
  {2004})}\BibitemShut {NoStop}%
\bibitem [{\citenamefont {Yarkony}\ and\ \citenamefont
  {Silbey}(1976)}]{Silbey_1976_comments}%
  \BibitemOpen
  \bibfield  {author} {\bibinfo {author} {\bibfnamefont {D.}~\bibnamefont
  {Yarkony}}\ and\ \bibinfo {author} {\bibfnamefont {R.}~\bibnamefont
  {Silbey}},\ }\href {\doibase 10.1063/1.433182} {\bibfield  {journal}
  {\bibinfo  {journal} {The Journal of Chemical Physics}\ }\textbf {\bibinfo
  {volume} {65}},\ \bibinfo {pages} {1042} (\bibinfo {year}
  {1976})}\BibitemShut {NoStop}%
\bibitem [{\citenamefont {Kvasnikov}(1958)}]{BI_1958}%
  \BibitemOpen
  \bibfield  {author} {\bibinfo {author} {\bibfnamefont {I.~A.}\ \bibnamefont
  {Kvasnikov}},\ }\href {http://mi.mathnet.ru/dan22873} {\bibfield  {journal}
  {\bibinfo  {journal} {Dokl. Akad. Nauk SSSR}\ }\textbf {\bibinfo {volume}
  {119}},\ \bibinfo {pages} {475} (\bibinfo {year} {1958})}\BibitemShut
  {NoStop}%
\bibitem [{\citenamefont {Predescu}(2002)}]{BI_2002}%
  \BibitemOpen
  \bibfield  {author} {\bibinfo {author} {\bibfnamefont {C.}~\bibnamefont
  {Predescu}},\ }\href@noop {} {\bibfield  {journal} {\bibinfo  {journal}
  {Physical Review E}\ }\textbf {\bibinfo {volume} {66}},\ \bibinfo {pages}
  {066133} (\bibinfo {year} {2002})}\BibitemShut {NoStop}%
\bibitem [{\citenamefont
  {Kuzemsky}(2015)}]{Kuzemsky_2015_Variational_principle_of_Bogoliubov}%
  \BibitemOpen
  \bibfield  {author} {\bibinfo {author} {\bibfnamefont {A.~L.}\ \bibnamefont
  {Kuzemsky}},\ }\href {\doibase 10.1142/s0217979215300108} {\bibfield
  {journal} {\bibinfo  {journal} {International Journal of Modern Physics B}\
  }\textbf {\bibinfo {volume} {29}},\ \bibinfo {pages} {1530010} (\bibinfo
  {year} {2015})}\BibitemShut {NoStop}%
\bibitem [{\citenamefont {Cheng}\ and\ \citenamefont
  {Silbey}(2008)}]{Silbey_2008_general_free_energy}%
  \BibitemOpen
  \bibfield  {author} {\bibinfo {author} {\bibfnamefont {Y.-C.}\ \bibnamefont
  {Cheng}}\ and\ \bibinfo {author} {\bibfnamefont {R.~J.}\ \bibnamefont
  {Silbey}},\ }\href {\doibase 10.1063/1.2894840} {\bibfield  {journal}
  {\bibinfo  {journal} {The Journal of Chemical Physics}\ }\textbf {\bibinfo
  {volume} {128}},\ \bibinfo {pages} {114713} (\bibinfo {year}
  {2008})}\BibitemShut {NoStop}%
\bibitem [{\citenamefont {Jaynes}\ and\ \citenamefont
  {Cummings}(1963)}]{jaynes1963comparison}%
  \BibitemOpen
  \bibfield  {author} {\bibinfo {author} {\bibfnamefont {E.~T.}\ \bibnamefont
  {Jaynes}}\ and\ \bibinfo {author} {\bibfnamefont {F.~W.}\ \bibnamefont
  {Cummings}},\ }\href@noop {} {\bibfield  {journal} {\bibinfo  {journal}
  {Proceedings of the IEEE}\ }\textbf {\bibinfo {volume} {51}},\ \bibinfo
  {pages} {89} (\bibinfo {year} {1963})}\BibitemShut {NoStop}%
\bibitem [{\citenamefont {Wu}\ \emph {et~al.}(2016)\citenamefont {Wu},
  \citenamefont {Feist},\ and\ \citenamefont
  {Garcia-Vidal}}]{Wu_2016_polarons_meet_polaritons}%
  \BibitemOpen
  \bibfield  {author} {\bibinfo {author} {\bibfnamefont {N.}~\bibnamefont
  {Wu}}, \bibinfo {author} {\bibfnamefont {J.}~\bibnamefont {Feist}}, \ and\
  \bibinfo {author} {\bibfnamefont {F.~J.}\ \bibnamefont {Garcia-Vidal}},\
  }\href {http://dx.doi.org/10.1103/PhysRevB.94.195409} {\bibfield  {journal}
  {\bibinfo  {journal} {Physical Review B}\ }\textbf {\bibinfo {volume} {94}},\
  \bibinfo {pages} {195409} (\bibinfo {year} {2016})}\BibitemShut {NoStop}%
\bibitem [{\citenamefont {Carmichael}(1999)}]{carmichaael}%
  \BibitemOpen
  \bibfield  {author} {\bibinfo {author} {\bibfnamefont {H.~J.}\ \bibnamefont
  {Carmichael}},\ }\enquote {\bibinfo {title} {Statistical methods in quantum
  optics 1},}\ \ (\bibinfo  {publisher} {Springer-Verlag Berlin Heidelberg},\
  \bibinfo {year} {1999})\ pp.\ \bibinfo {pages} {22--24}\BibitemShut {NoStop}%
\bibitem [{\citenamefont {Roy-Choudhury}\ and\ \citenamefont
  {Hughes}(2015)}]{Hughes_2015_dipole_emission}%
  \BibitemOpen
  \bibfield  {author} {\bibinfo {author} {\bibfnamefont {K.}~\bibnamefont
  {Roy-Choudhury}}\ and\ \bibinfo {author} {\bibfnamefont {S.}~\bibnamefont
  {Hughes}},\ }\href {https://link.aps.org/doi/10.1103/PhysRevB.92.205406}
  {\bibfield  {journal} {\bibinfo  {journal} {Phys. Rev. B}\ }\textbf {\bibinfo
  {volume} {92}},\ \bibinfo {pages} {205406} (\bibinfo {year}
  {2015})}\BibitemShut {NoStop}%
\bibitem [{\citenamefont {Cosacchi}\ \emph {et~al.}(2021)\citenamefont
  {Cosacchi}, \citenamefont {Seidelmann}, \citenamefont {Cygorek},
  \citenamefont {Vagov}, \citenamefont {Reiter},\ and\ \citenamefont
  {Axt}}]{cosacchi2021accuracy}%
  \BibitemOpen
  \bibfield  {author} {\bibinfo {author} {\bibfnamefont {M.}~\bibnamefont
  {Cosacchi}}, \bibinfo {author} {\bibfnamefont {T.}~\bibnamefont
  {Seidelmann}}, \bibinfo {author} {\bibfnamefont {M.}~\bibnamefont {Cygorek}},
  \bibinfo {author} {\bibfnamefont {A.}~\bibnamefont {Vagov}}, \bibinfo
  {author} {\bibfnamefont {D.~E.}\ \bibnamefont {Reiter}}, \ and\ \bibinfo
  {author} {\bibfnamefont {V.~M.}\ \bibnamefont {Axt}},\ }\href@noop {}
  {\bibfield  {journal} {\bibinfo  {journal} {ArXiv preprint ArXiv:2103.13100}\
  } (\bibinfo {year} {2021})}\BibitemShut {NoStop}%
\bibitem [{\citenamefont {Kaer}\ and\ \citenamefont
  {M\o{}rk}(2014)}]{Kaer_2014_cav_emission}%
  \BibitemOpen
  \bibfield  {author} {\bibinfo {author} {\bibfnamefont {P.}~\bibnamefont
  {Kaer}}\ and\ \bibinfo {author} {\bibfnamefont {J.}~\bibnamefont {M\o{}rk}},\
  }\href {\doibase 10.1103/PhysRevB.90.035312} {\bibfield  {journal} {\bibinfo
  {journal} {Phys. Rev. B}\ }\textbf {\bibinfo {volume} {90}},\ \bibinfo
  {pages} {035312} (\bibinfo {year} {2014})}\BibitemShut {NoStop}%
\bibitem [{\citenamefont {Hornecker}\ \emph
  {et~al.}(2017{\natexlab{b}})\citenamefont {Hornecker}, \citenamefont
  {Auff\`eves},\ and\ \citenamefont {Grange}}]{Hornecker_2017_cav_emission}%
  \BibitemOpen
  \bibfield  {author} {\bibinfo {author} {\bibfnamefont {G.}~\bibnamefont
  {Hornecker}}, \bibinfo {author} {\bibfnamefont {A.}~\bibnamefont
  {Auff\`eves}}, \ and\ \bibinfo {author} {\bibfnamefont {T.}~\bibnamefont
  {Grange}},\ }\href {\doibase 10.1103/PhysRevB.95.035404} {\bibfield
  {journal} {\bibinfo  {journal} {Phys. Rev. B}\ }\textbf {\bibinfo {volume}
  {95}},\ \bibinfo {pages} {035404} (\bibinfo {year}
  {2017}{\natexlab{b}})}\BibitemShut {NoStop}%
\bibitem [{\citenamefont {Faraon}\ \emph {et~al.}(2012)\citenamefont {Faraon},
  \citenamefont {Santori}, \citenamefont {Huang}, \citenamefont {Acosta},\ and\
  \citenamefont {Beausoleil}}]{NV_center1}%
  \BibitemOpen
  \bibfield  {author} {\bibinfo {author} {\bibfnamefont {A.}~\bibnamefont
  {Faraon}}, \bibinfo {author} {\bibfnamefont {C.}~\bibnamefont {Santori}},
  \bibinfo {author} {\bibfnamefont {Z.}~\bibnamefont {Huang}}, \bibinfo
  {author} {\bibfnamefont {V.~M.}\ \bibnamefont {Acosta}}, \ and\ \bibinfo
  {author} {\bibfnamefont {R.~G.}\ \bibnamefont {Beausoleil}},\ }\href@noop {}
  {\bibfield  {journal} {\bibinfo  {journal} {Phys. Rev. Lett.}\ }\textbf
  {\bibinfo {volume} {109}},\ \bibinfo {pages} {033604} (\bibinfo {year}
  {2012})}\BibitemShut {NoStop}%
\bibitem [{\citenamefont {McCutcheon}\ and\ \citenamefont
  {Lon\v{c}ar}(2008)}]{NV_center2}%
  \BibitemOpen
  \bibfield  {author} {\bibinfo {author} {\bibfnamefont {M.~W.}\ \bibnamefont
  {McCutcheon}}\ and\ \bibinfo {author} {\bibfnamefont {M.}~\bibnamefont
  {Lon\v{c}ar}},\ }\href
  {http://www.opticsexpress.org/abstract.cfm?URI=oe-16-23-19136} {\bibfield
  {journal} {\bibinfo  {journal} {Opt. Express}\ }\textbf {\bibinfo {volume}
  {16}},\ \bibinfo {pages} {19136} (\bibinfo {year} {2008})}\BibitemShut
  {NoStop}%
\bibitem [{\citenamefont {Wigger}\ \emph {et~al.}(2020)\citenamefont {Wigger},
  \citenamefont {Karakhanyan}, \citenamefont {Schneider}, \citenamefont {Kamp},
  \citenamefont {Höfling}, \citenamefont {Machnikowski}, \citenamefont
  {Kuhn},\ and\ \citenamefont {Kasprzak}}]{Wigger_2020_polaron_sideband}%
  \BibitemOpen
  \bibfield  {author} {\bibinfo {author} {\bibfnamefont {D.}~\bibnamefont
  {Wigger}}, \bibinfo {author} {\bibfnamefont {V.}~\bibnamefont {Karakhanyan}},
  \bibinfo {author} {\bibfnamefont {C.}~\bibnamefont {Schneider}}, \bibinfo
  {author} {\bibfnamefont {M.}~\bibnamefont {Kamp}}, \bibinfo {author}
  {\bibfnamefont {S.}~\bibnamefont {Höfling}}, \bibinfo {author}
  {\bibfnamefont {P.}~\bibnamefont {Machnikowski}}, \bibinfo {author}
  {\bibfnamefont {T.}~\bibnamefont {Kuhn}}, \ and\ \bibinfo {author}
  {\bibfnamefont {J.}~\bibnamefont {Kasprzak}},\ }\href {\doibase
  10.1364/ol.385602} {\bibfield  {journal} {\bibinfo  {journal} {Optics
  Letters}\ }\textbf {\bibinfo {volume} {45}},\ \bibinfo {pages} {919}
  (\bibinfo {year} {2020})}\BibitemShut {NoStop}%
\bibitem [{\citenamefont {McCutcheon}(2010)}]{dara_phd}%
  \BibitemOpen
  \bibfield  {author} {\bibinfo {author} {\bibfnamefont {D.~P.~S.}\
  \bibnamefont {McCutcheon}},\ }\emph {\bibinfo {title} {Open quantum systems
  in spatially correlated regimes}},\ \href@noop {} {Ph.D. thesis},\ \bibinfo
  {school} {University College London} (\bibinfo {year} {2010})\BibitemShut
  {NoStop}%
\end{thebibliography}
\end{document}